\documentclass[useAMS,usenatbib]{mn2e}
\usepackage{graphicx}
\usepackage{times}
\usepackage{amssymb,amsmath}
\usepackage{natbib}
\usepackage{color}
\usepackage{fixltx2e}

\def\gsim { \lower .75ex \hbox{$\sim$} \llap{\raise .27ex \hbox{$>$}} }
\def\lsim { \lower .75ex \hbox{$\sim$} \llap{\raise .27ex \hbox{$<$}} }

\newcommand{\e}[2][def]{
\begin{equation}
#2 \label{#1}
\end{equation} }

\usepackage{latexsym,amsmath,amssymb,amsfonts,enumerate}
\usepackage[mathscr]{eucal}
\usepackage{makeidx}
\usepackage{epsfig}

\usepackage{ifthen}
\usepackage{graphicx}

\newcommand{\MHz}{{~\rm{MHz}}}

\title[The LOFAR Data Model]{The LOFAR EoR Data Model: (I) Effects of Noise and Instrumental Corruptions
on the 21-cm Reionization Signal-Extraction Strategy}

\author[Labropoulos et al. ]{P. Labropoulos$^{1}$\thanks{E-mail:
panos@astro.rug.nl}, L. V. E. Koopmans$^{1}$, V. Jeli{\'c}$^{1}$, S. Yatawatta$^{1}$,  R. M. Thomas$^{1}$, 
\newauthor G. Bernardi$^{1}$,  M. Brentjens$^{2}$,
A.G. de Bruyn$^{1,2}$,  B. Ciardi$^{3}$, G. Harker$^{1}$, A.R. Offringa$^{1}$, \newauthor V. N. Pandey$^{1}$,  J.  Schaye$^{4}$,
S. Zaroubi$^{1}$\\~\\ $^{1}$Kapteyn Astronomical Institute, University
of Groningen, P.O. Box 800, 9700 AV Groningen, the Netherlands\\
$^{2}$ASTRON, Postbus 2, 7990 AA Dwingeloo, the Netherlands\\
$^{3}$Max-Planck Institute for Astrophysics,
Karl-Schwarzschild-Stra\ss e 1, 85748 Garching, Germany\\ $^{4}$Leiden
Observatory, Leiden University, PO Box 9513, 2300 RA Leiden, the
Netherlands}

\begin{document}


\pagerange{\pageref{firstpage}--\pageref{lastpage}} \pubyear{2008}

\maketitle

\label{firstpage}

\begin{abstract}

A number of experiments are set to measure the 21-cm signal of neutral hydrogen from the Epoch of 
Reionization (EoR).
The common denominator of these experiments are the large data sets produced, contaminated by 
various instrumental effects, ionospheric distortions, RFI and strong Galactic and extragalactic 
foregrounds. In this paper, the first in a series, we present the {\sl Data Model} that will be the basis of the 
signal analysis for the LOFAR (Low Frequency Array) EoR Key Science Project (LOFAR EoR KSP). 
Using this data model we simulate realistic visibility data sets over a wide frequency band, taking properly into 
account all currently known instrumental corruptions (e.g. direction-dependent gains, complex gains, polarization 
effects, noise, etc).  
We then apply primary calibration errors to the data in a statistical sense, assuming that the calibration errors
are random Gaussian variates at a level consistent with our current knowledge based on observations
with the LOFAR Core \hbox{Station 1}.  Our aim is to demonstrate how 
the systematics of an interferometric measurement affect the quality of the calibrated data, how errors 
correlate and propagate, and in the long run how this can lead 
to new calibration strategies.  We present results of these simulations and the 
inversion process and extraction procedure. We also discuss some general properties of the coherency matrix
and Jones formalism that might prove useful in solving the calibration problem of aperture
synthesis arrays. We conclude that even in the presence of realistic  noise and instrumental errors, the statistical 
signature of the EoR signal can be detected by LOFAR with relatively small errors. A detailed study of the statistical properties of 
our data model and more complex instrumental models will be considered in the future.
 
 \end{abstract}


\section{Introduction}

Recent years have seen a marked increase in the study, both
theoretical and observational, of the epoch in the history of our
Universe after the cosmological recombination era: from the so called
\textit{`Dark Ages'} to the \textit{Epoch of Reionization} (EoR)
\citep{hogan79,scott90,madau97}. A cold and dark Universe, after the
recombination era, was illuminated by sources of radiation, be it
stars, quasars or dark matter annihilation. These `first objects'
ionized and heated their surrounding inter-galactic medium (IGM),
carving out `bubbles' in the otherwise neutral hydrogen-filled
Universe. These bubbles grew rapidly, both in size and number, and
caused a phase transition in the hydrogen-ionized fraction of our
Universe at redshifts 6$<$$z$$<$20 \citep{sz75}. Although the EoR
spanned a relatively small fraction, in time, of the Universe's age,
its impact on subsequent structure formation (at least baryonic) is
crucial. Hence, studying the EoR directly influences our understanding
of issues in contemporary astrophysical research such as metal-poor
stars, early galaxy formation, quasars and cosmology
\citep{nusser05,zaroubi05,kuhlen05,thomas08,field58, field59, scott90,
kumar95, madau97}. For a detailed review of the EoR and our current
efforts to detect it, we refer the reader to \citet{FOB} and the
references therein.

Given the recent progress in developing a concrete theoretical
framework, and simulations based thereon, the EoR from an
observational point of view is still very poorly constrained. Despite
a wealth of observational cosmological data made available during the
past years \citep[e.g.~][]{spergel07, page07,becker01, fan01,
pentericci02, white03, fan06}, data directly probing the EoR have
eluded scientists and the ones that constrain the EoR are indirect and
very model-dependent \citep{barkana01, loeb01, ciardi03a, ciardi03b,
bromm04, iliev07, zaroubi07, thomas08}. Currently there are two main
observational constraints on the EoR: first,
the sudden jump in the Lyman-$\alpha$ optical depth in the
Gunn--Peterson troughs \citep{gunn65}, observed in the quasar spectra
of the Sloan Digital Sky Survey (SDSS) (Becker et al. 2001; Fan et
al. 2001; Pentericci et al. 2002; White et al.q 2003; Fan et al. 2006)
which provides a limit on when reionization was completed. Current
consensus is that reionization ended around a redshift of six. Second,
the five-year \emph{WMAP} data on the temperature and polarization
anisotropies of the cosmic microwave background (CMB) (Spergel et
al. 2007; Page et al. 2007) which gives an integral constraint on the
Thomson optical depth for scattering experienced by the CMB photons
since the EoR. A maximum likelihood analysis performed by
\citet{spergel07} estimates the peak of reionization to have occured at 11.3
when the cosmic age was 365 Myr. Thus, we see that current
astronomical data is only able to provide us with crude boundaries within
which reionization occured. In order to properly characterize the
onset, evolution and completion of the EoR and derive results on its
impact on subsequent evolution of structures in the Universe, we need
more direct measurements from the EoR.

\begin{figure*}
  \centering
\includegraphics[width=0.27\textwidth]{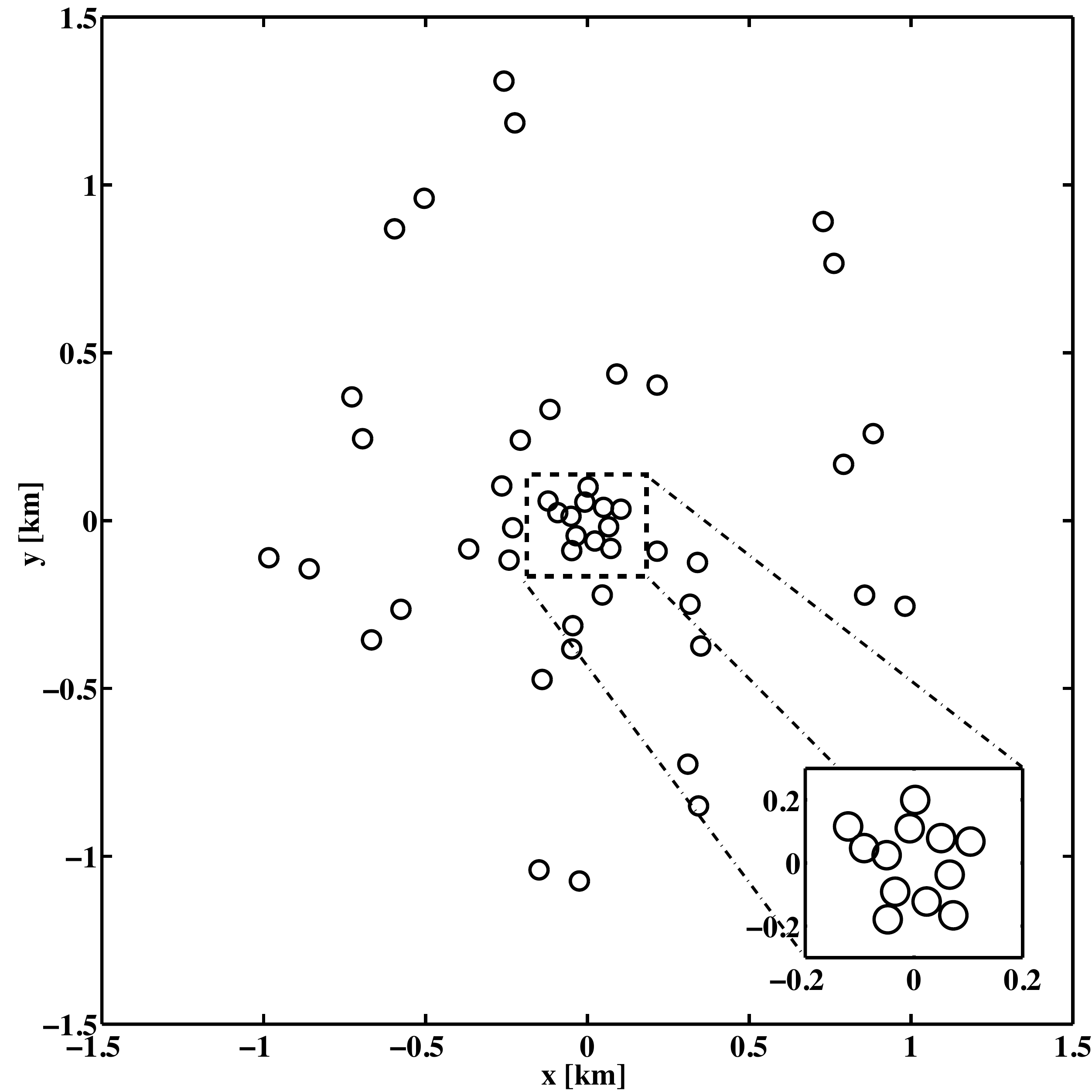} 
\includegraphics[width=0.27\textwidth]{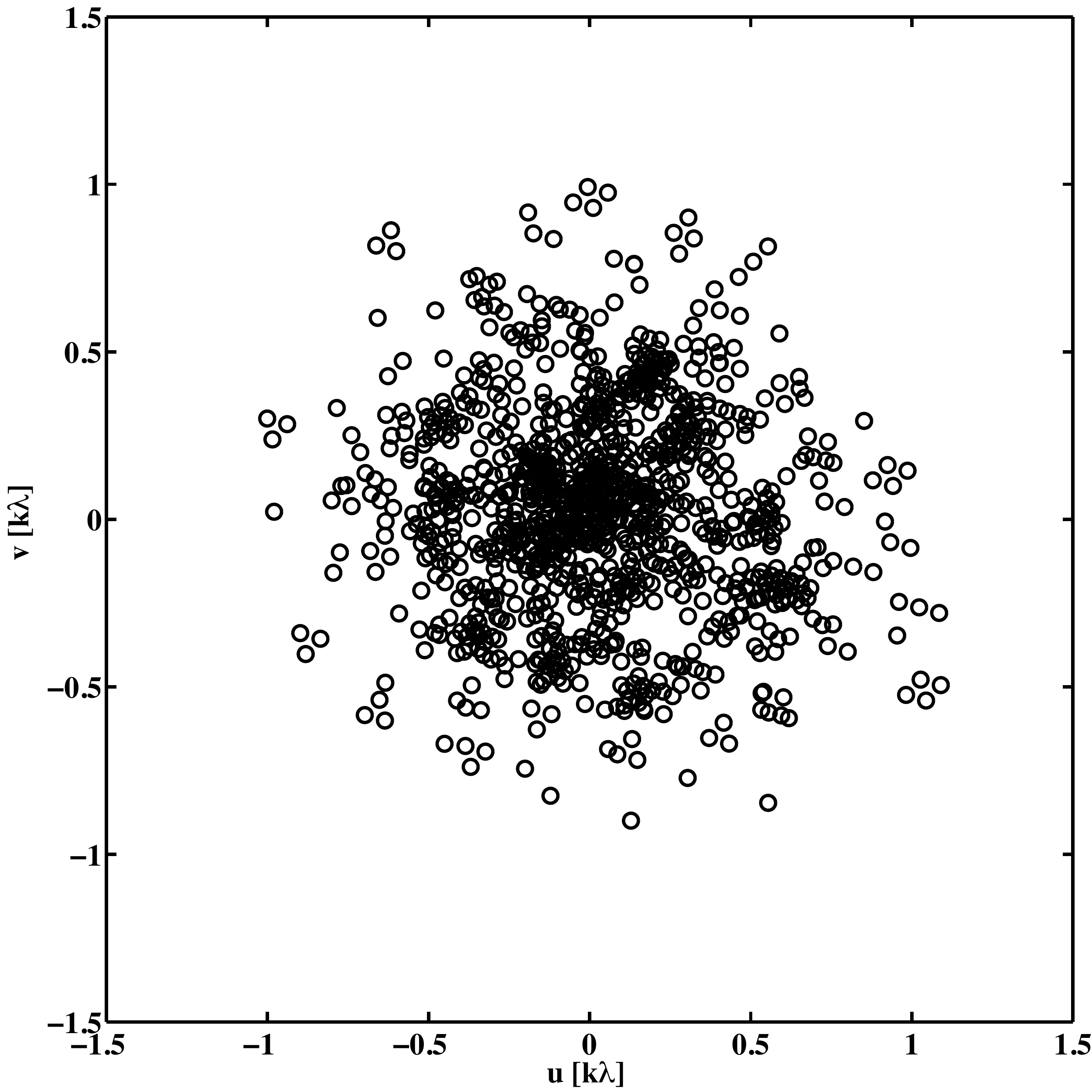} 
\includegraphics[width=0.37\textwidth]{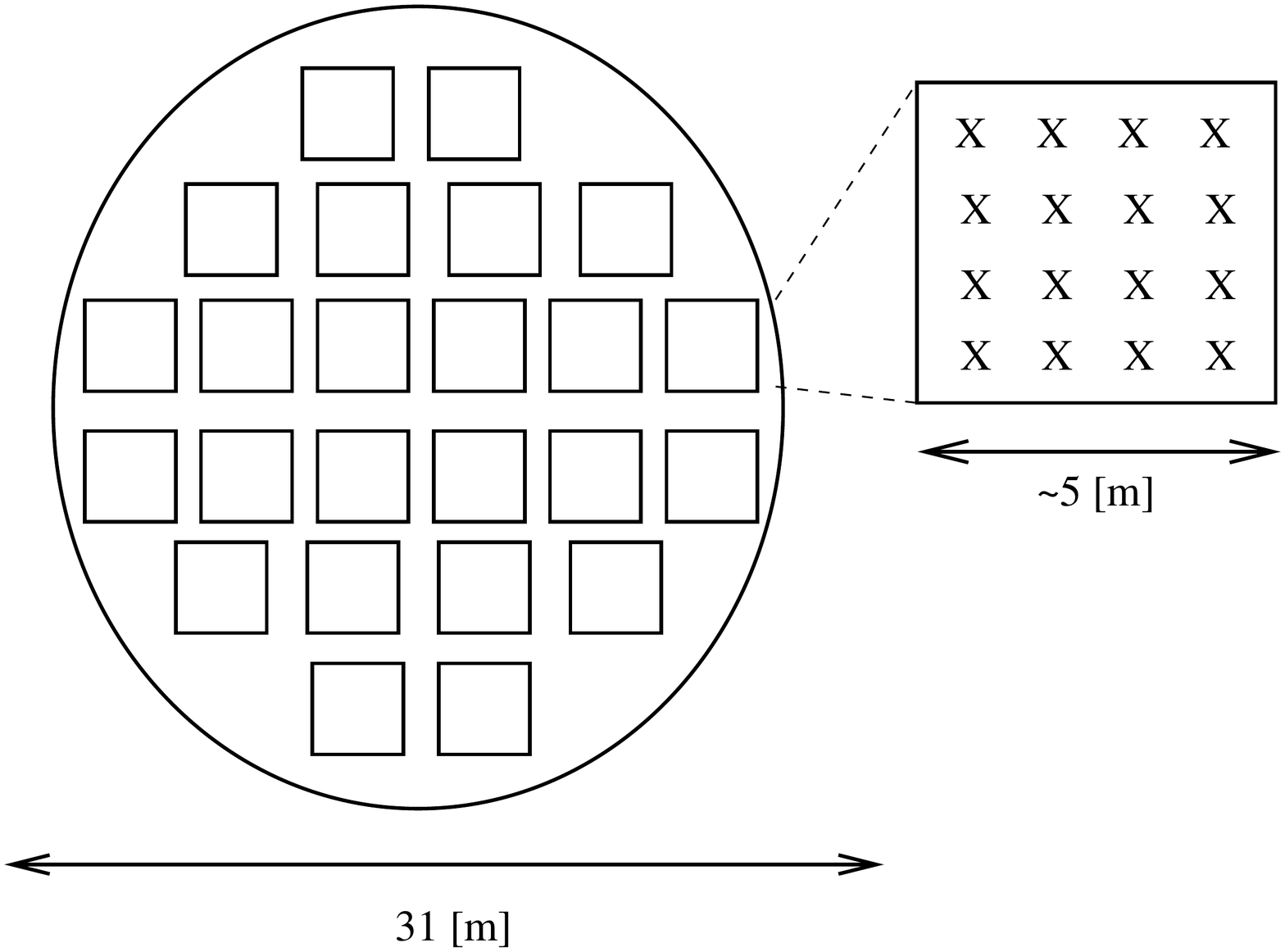}

\label{fig:layout}\label{fig:HBA_tile}
\caption{{\bf Left:} The LOFAR compact-core layout situated near Exloo
in the Netherlands. The open circle-size corresponds to the High Band
Antenna (HBA) station size of approximately 31 metres in diameter. The
6 inner stations (``six--pack'') are shown in the the inset
figure. {\bf Middle:} The snapshot uv-coverage of the LOFAR compact
core at the zenith and at a frequency 150 ~MHZ. {\bf Right:} The layout of a LOFAR HBA station. Each square represents a tile that consists of four by four, 
orthogonal bowtie dipoles, as shown in the blown--up inset. Each dipole pair can be rotated individually during the installation of the tile.  }
\end{figure*}

Observations of the hydrogen 21-cm hyperfine ``spin-flip'' transition,
using radio interferometry, provide just such a direct
probe of the dark ages and the EoR over a wide spatial and redshift range. It is
worth mentioning that the ``spatial range'' here implies the two
dimensions on the sky, which is a function of the baseline lengths of
an interferometer, and the third dimension along the redshift
direction, which depends on the frequency resolution of the
observation. The 21-cm emission line from the EoR is redshifted by
$1+z$, because of the expansion of the Universe, to wavelengths in the
meter waveband. For example, at a redshift of $z=9$ the 21-cm line is
redshifted to 2.1 metres, which corresponds to a frequency of about
140 \MHz. Computer simulations suggest that we may expect a complex, evolving
patch-work of neutral (H$_\mathrm{I}$) and ionized hydrogen (H$_\mathrm{II}$)
regions. If we manage to successfully image the Universe at these high
redshifts ($6 < z < 12$) we expect to find H$_\mathrm{II}$ regions, created by
ionizing radiation from first objects, to appear as ``holes'' in an
otherwise neutral hydrogen-filled Universe; the so-called
\emph{Swiss-cheese model}. Current constraints and simulations
converge on reionization happening for a large part in the redshift range $z\approx11.4$ ($\sim$115
\MHz) to $z\approx6$ ($\sim$203 \MHz), which is the range probed by LOFAR\footnote{Low Frequency Array: http://www.lofar.org}, a radio 
interferometer currently being built near the village of  Exloo in the Netherlands. The
21-cm radiation can not only trace the matter power spectrum in the
period after recombination, but also can constrain reionization
scenarios \citep{thomas08,barkana01}. Note that because strongly
radiating sources create bubbles of ionized gas inside the neutral
IGM, one should observe fluctuations in the 21-cm emission due to
reionization that deviate from those of neutral gas tracing the
dark-matter distribution, even deep into the highly linear regime.

Developments in radio-wave sensor technologies in recent years have
enabled us to conceive of and design extremely large, high sensitivity
and high resolution radio interferometers, a development which is essential to
conduct a successful 21-cm experiment to image the EoR. A series
of radio telescopes are being built similar to LOFAR, 
such as MWA\footnote{Murchison Widefield Array: http://www.haystack.mit.edu/ast/arrays/mwa}, PAPER
\footnote{Precision Array to Probe Epoch of Reionization: http://astro.berkeley.edu/~dbacker/eor/},
21CMA\footnote{21-cm Array: http://web.phys.cmu.edu/~past/} and further in the
future the SKA\footnote{Square Kilometer Array: http://www.skatelescope.org}, all with one
of their primary goals being the detection of  the
redshifted 21-cm signal from the
EoR. The GMRT\footnote{Giant Meterwave Radio Telescope: http://www.gmrt.ncra.tifr.res.in} has a programme already
under way to detect the EoR or at least to constrain the foregrounds
that may hamper the experiment \citep{pen08}.

Calculations predict the cosmological 21-cm signal from the EoR to be extremely
faint. Apart from the intrinsic low strength of the 21-cm
signal, the experiment is plagued by a myriad of signal contaminants like
man-made and natural (e.g. lightnings) interference, ionospheric distortions, Galactic free-free and
synchrotron radiation, clusters and radio galaxies along the path
of the signal. Thus long integration times, exquisite calibration and
well-designed RFI mitigation techniques are needed in order to ensure the
detection of the underlying signal. It is also imperative to properly model all these effects beforehand, in order to develop
sophisticated schemes that will be needed to clean the data cubes from
these contaminants
\citep{shaver99,dimatteo04,dimatteo02,oh03,cooray04,zaldarriaga04,
gleser07}. Due to the low signal-to-noise ratio per resolution element
(of the order of 0.2 or for LOFAR and and even less for e.g. MWA),
the initial aim of all current experiments is to obtain a statistical
detection of the signal. By statistical, we mean a global change in
a property of the signal, for example the variance, as a function of
frequency and angular scale. Note that this task involves distinguishing these
statistical properties from those of the calibration residuals and the
thermal noise.

In addition to the above-mentioned astrophysical and terrestrial sources of contamination, one also has  to face 
issues arising in standard synthesis imaging. For that it is crucial to describe all physical effects on the signal that determine the values of the measured visibilities. The study of 
polarized radiation falls within the regime of optics: \cite{hbs1} provided such a unifying model for the Jones and Mueller 
calculi in optics \citep{bornwolf} and the techniques of radio interferometry based on multiplying correlators. Because low
frequency phased-array dipole antennas are inherently polarized, one has to consider polarimetry from the 
beginning. The Measurement Equation (ME) of \citet{hbs1}  is therefore a natural way to describe LOFAR. Their 
treatment \citep{hbs1,hbs3,hbs4,hbs5} forms the basis of our data model description and we will  present it in a wider setting, giving the 
connection to physics. The Hamaker--Bregman--Sault measurement equation acts on the astronomical signals as a ``black-box'': the 
interferometer converts the input signal Stokes vectors to the final output at the correlator. This is done via a sequence 
of linear transformations and thus enables us to systematically model the series of effects that modify the signal while it propagates through the ionosphere and the receivers. 

Calibration \citep{miguel08a} of the observed visibility data set is generally aimed at determining instrumental 
parameters of the antennas at a level sufficient to detect signals at several times the noise level. 
While this is the traditional 
approach, a much more thorough understanding of the instrumental response is required in currently designed 
experiments such as the LOFAR EoR Key-Science Project, where unprecedented high dynamic range and 
sensitivity have to be achieved, and the signal is far below the noise level. 
For example, the CLEAN algorithm and variations on it have been used extensively, as an integral part of the SELFCAL process \citep{selfcal1}.
While computationally efficient, it does not provide a 
statistically optimal solution \citep{schwarz78,starck02}. In this work we shall 
therefore consider a maximum-likelihood solution to the 
measurement equation inversion problem \citep{boonstra05,vdv00a,vdv00b,stefan04}, which takes 
into account realistic zero-mean calibration residuals and noise. We are also examining alternative solutions \citep{sarod08b}, however.

After giving a short description of the LOFAR array, with emphasis on
the high band (HBA) aspects of the design in Section 2, we review
the basic relation between the observed visibilities and the sky
intensity in section 3. We emphasize the polarized, matrix formulation
of the measurement equation and the mathematical aspects of coherency
matrix \citep{hbs1}. In Section 4 we briefly discuss the relevant
Measurement Equation (ME) parameters for LOFAR. Using this measurement
equation as our data model, in Section 5, we produce a number of
simulations for different instrumental parameters. This is the forward
use of the data model. We also try to invert the data model, given the
instrumental parameters and their error distributions, in order to
recover the original data. Our goal is to test the calibration and
inversion requirements using realistically generated data cubes. In
Section 6 we discuss the results and give our conclusions.

\section{Description of the Low-Frequency Array}

The immediate science goals of the LOFAR EoR-KSP, that drive some of the considerations about 
the design of LOFAR \citep{lofref}, are: 
(1) extract the 21-cm neutral hydrogen signal averaged 
along lines of sight, i.e.\ the `global signal' \citep[e.g.][]{shaver99, jelic08}, (2) determine the spatial-frequency power 
spectrum of the brightness temperature fluctuations on angular scales of  about 1 arcminute to 1 degree and 
frequency scales between 0.1
and 10 MHz in the redshift range of $\sim$6-11 and (3) search for Str\"omgren ionization bubbles around bright sources and the 21-cm absorption-line forest
\citep{PP07}. 
In order to achieve these science goals, LOFAR requires a good uv-coverage, a good frequency coverage and a 
large collecting area.	

Below, we give a short summary of the aspects of LOFAR which are relevant to the EoR experiment and 
our data model. For more details we refer to the project paper by de Bruyn et al. (in preparation).

\subsection{Station configuration and uv-coverage}	

\begin{figure*}
	\centering
		\includegraphics[width=1.1\textwidth]{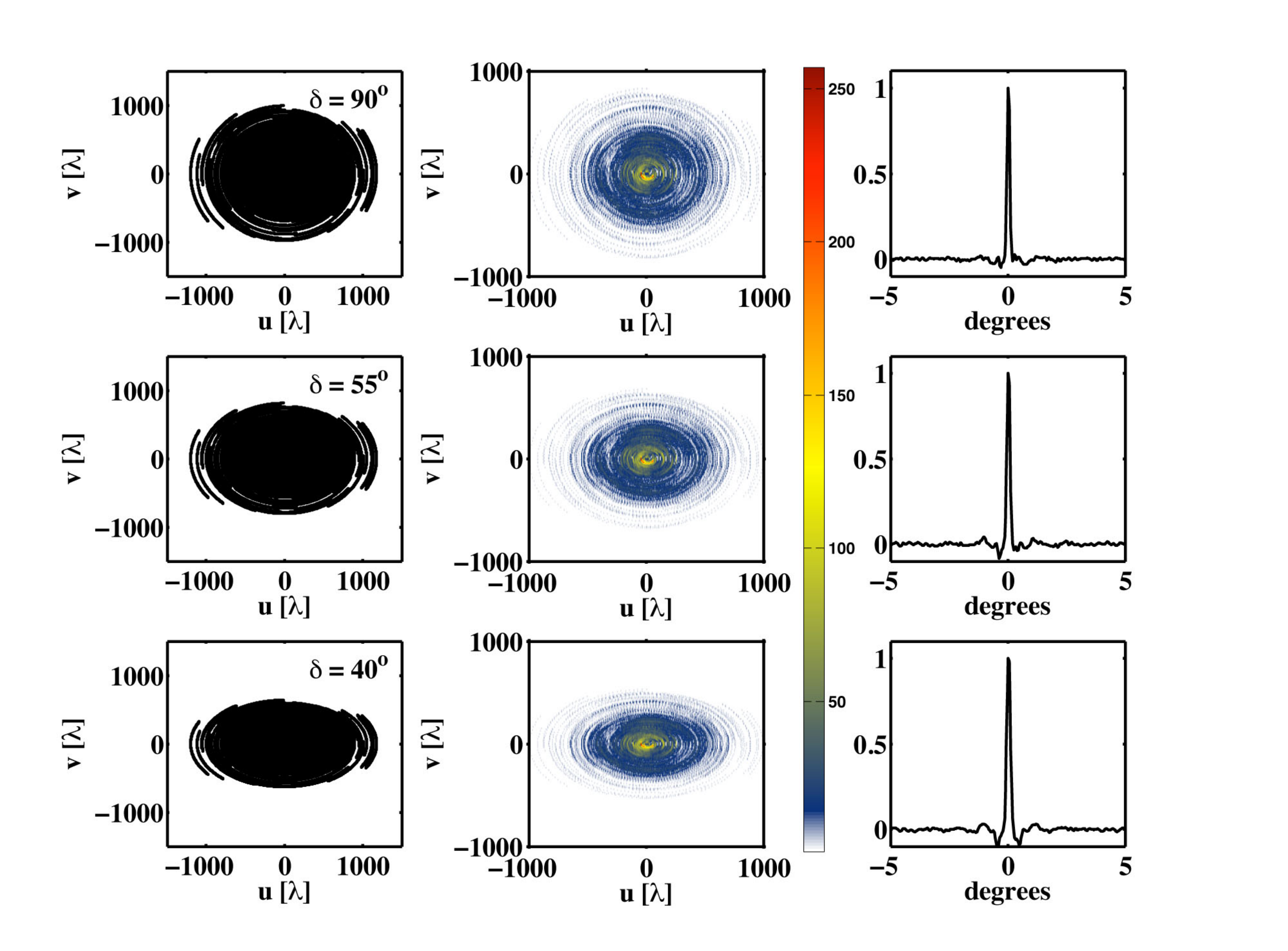}
	
\caption{The first column of figures shows the uv-coverage of the LOFAR core, that will be used for the EOR experiment, for diffrent values of the declination $\delta$. The size of each point corresponds to the station diameter. 
The uv coverage is calculated for 4 hours of synthesis with 10s averaging at 150 \MHz. The second column 
shows the corresponding uv point density. The uv plane is gridded with a cell size of 8.5 wavelengths squared. 
In the last column a horizontal cut of the ``dirty'' beam is shown.}
\label{fig:plot9x9}
\end{figure*}

In its current layout,  the LOFAR telescope \citep{PP07,falcke07} will consist of up to 48 stations of 
which approximately 24 will be located in the core region (Figure~\ref{fig:layout}), near the village of Exloo in the
Netherlands. The core marks an area of 1.7 by 2.3 kilometres.  Each High Band Antenna station (HBA station; 110--240 MHz;
see next section) in the core is further split into two ``half-stations'' of half the collecting area 
($\sim$31 metre 
diameter), separated by $\sim$130~metres. This split further improves the uv-coverage, though  at the cost of quadrupling the BlueGene/P 
correlator
demands (i.e. there are four times more base-lines). The central region of the core consists of six closely-packed 
stations, ``the six-pack'', to ensure improved coverage of the shortest baselines necessary to map
out the largest scales on the sky, such as the Milky Way.
The station-layout yields a snapshot uv-coverage at the zenith as shown in Figure ~\ref{fig:layout}. The uv-coverages for a typical synthesis time of 4 \rm{hrs} are shown in Figure ~\ref{fig:plot9x9}.

A good uv-coverage is crucial for several reasons. First to improve sampling of the power spectrum of the EoR 
signal
\citep{santos05,hobson02,miguel08,miguel06,miguel05a,miguel05b}. Second, to obtain precise Local \citep{lsm} and Global \citep{gsm} Sky models (LSM/GSM; i.e.\ catalogues of the brightest, 
mostly compact, sources in and outside of the beam, i.e. local versus global). A further complication is the extraction of the 
Galactic and extragalactic foregrounds. This is a vital step in the recovery of the signal and requires good 
sampling of the uv-plane at all frequencies \citep{miguel08,miguel06,miguel05b}.

\subsection{The High-Band Antennas}

LOFAR will have two sets of dipoles, the Low Band Antennas (LBA) and the High Band Antennas (HBA). For the 
EoR experiment we are mostly interested in the HBA dipoles which cover the 110 to 220 MHz frequency range. Each dipole is a crossed dipole which 
enables X and Y polarization observations. Inside the station the dipoles are arranged in tiles of 4 by 4 dipoles, 
with 24 tiles per station inside the core (Figure ~\ref{fig:layout}). 
Radio waves are sampled with a 16-bit analog-to-digital converter to be able to cope with expected interference levels 
operating at either 160 or 200 MHz in the first, second or third Nyquist zone (i.e., 0--100, 100--200, or 200--300 
MHz band respectively for 200 MHz sampling). The data from
the receptors are filtered in 512, 195 kHz sub-bands (156 kHz sub-bands for 160 MHz
sampling) of which a total of 32 MHz bandwidth (164 channels) can be used at one time. Sub-bands from all 
antennas are combined at the  station level in
a digital beamformer allowing multiple (4--6) independently steerable beams, which are sent to the
central processor via a glass-fibre link that handles 0.7 Tbit/s data. The beams from all
stations are further filtered into 1 kHz channels, cross-correlated and integrated. The integrated visibilities are then calibrated on 1 second
 intervals, to correct for the effects of the ionosphere, and 
subsequently images are produced. Channels
with disturbing radio frequency interference (RFI) are excised \citep{vdv00a,veen04,stefan04,fridman01}. For the correlation we
use three racks of an IBM Blue Gene/P machine in Groningen with a total of
12288 processing cores. LOFAR is a new concept in array design, a broad-band aperture array with digital
beamforming. This makes LOFAR essentially qualify as a pathfinder for the Square Kilometre Array \citep{falcke07}.

\section{Mathematical Framework of the LOFAR Data Model}

The most important part of any physical measurement is to find a correspondence between the physical 
quantities and the measured quantities. In radio interferometry this is achieved through the so called  
measurement equations
\citep{hbs1,boonstra05}. The measurement equation (ME) describes the relationship between the visibilities (correlations between the 
electric fields from different antennas) and the brightness distribution of the sky. We will begin by discussing 
briefly the different types of measurement equations and the implications for LOFAR \citep{noordam06,noordam04,noordam00}. 
The data model presented in this paper can be easily applied to other telescopes that operate at low 
frequencies such as the MWA and eventually the SKA.

Astronomical radio signals appear as spatially wide-band random noise with superimposed features, such as 
polarization, emission and absorption lines. The physical quantity that underlies this kind of measurement is the 
electric field, but for convenience astronomers try to recover the intensity in the direction of the unit pointing  
vector \textbf{s}, $I_f (\textbf{s}) = \left\langle  |E_{f} (\textbf{s})| \right\rangle ^2     $. The measured 
correlation of the electric fields between two sensors $i$ and $j$ is called the complex visibility. For Earth-rotation 
synthesis we assume that the telescopes have a small field of view (\rm{FOV}) and that they track a position on the sky. To 
achieve that, a slowly time-varying phase delay has to be introduced at the receiver to compensate for the 
geometrical delays. The result is that the reference location appears to be at the zenith, or their chosen point in the sky 
(phase reference center).
For a planar array, the receiver baselines can be parametrized as 
\e{
	\textbf{r}_i-\textbf{r}_j=\lambda[u,v,0],\phantom{dsdsf}\lambda\equiv\dfrac{c}{2\pi \nu}.
}
where $\bf{r}_i$ are the station position vectors. This system is wavelength  dependent. The (scalar) measurement equation in $(u,v)$ coordinates becomes
\e{
	V_f \left( u,v\right)   =\iint {\cal P}_f(l,m) I_f(l,m)\mathrm{e}^{-i\, (ul+lm)} \mathrm{d}l \mathrm{d}m
}
where ${\cal P}_f(l,m)$ is the complex primary beam or antenna response pattern and $I_f(l,m)$ is the sky brightness 
distribution.
This equation (van Cittert--Zernike theorem) is in the form of a 2--D Fourier transform, which is an approximation for a flat sky \citep{tms,woan}. The 
visibilities are sampled for all different sensor pairs $i$ and $j$, but also for different sensor locations projected 
on the sky, since the Earth rotates. Hence Earth-rotation synthesis traces uv tracks for each baseline \citep{tms}. 

\subsection{The Scalar Measurement Equation}

The most widely used ME is still the scalar formulation of the ME. We begin the 
discussion with the scalar ME and later we will also discuss the polarized version thereof. Since all processing is done with 
digital computers this equation must be transformed into a more convenient discretized form. The main output of 
the LOFAR correlator is a set of correlation matrices \citep{boonstra05,falcke07}, $
{\bf{R}}_f( {t_k })$, for a set of narrow-band frequency channels (adding up to 32 MHz bandwidth) and for a set of 
short-time integrations (adding
up to $>$300 hours of integration).  
The connection between the correlation matrices $
{\bf{R}}_f \left( {t_k } \right)$ and the visibilities $V_f \left( u,v \right)$ is that each entry $R_{ij}\left( t_k \right)$ of $
{\bf{R}}_f \left( {t_k } \right)$ is a sample of the visibility function for a specific coordinate $(u,v)$ corresponding to 
the baseline vector ${\bf{r}}_{ij}  = {\bf{r}}_i  - {\bf{r}}_j $ between telescopes $i$ and $j$ at time $t_k$ 
\citep{boonstra05}:
\[
V_f \left( {u_{ik}  - u_{jk} ,v_{ik}  - v_{jk} } \right) \equiv R_{ij} \left( {t_k } \right).
\]
The noiseless scalar measurement equation (not accounting for instrumental and other distorting effects) for one 
short-time integration and narrow-band frequency channel, assuming the sky can be described by a set of point-
sources, can then be written in terms of the correlation matrices\footnote{The symbol "$\dag $" stands for the 
Hermitian conjugation operator} as \citep{boonstra05,vdv00b}
\begin{equation}
	{\bf{R}}_{k,f}  = {\bf{A}}_{k,f} {\bf{B}}_f {\bf{A}}_{k,f}^\dag 
\end{equation}
where
\e{
\begin{array}{*{20}c}
   {{\bf{A}}_{k,f}  = \left[ {{\bf{a}}_{k,f} ({\bf{s}}_1 ), \ldots ,{\bf{a}}_{k,f} ({\bf{s}}_d )} \right]}  \\
  ~\\
   {{\bf{a}}_{k,f} ({\bf{s}}_i ) = \left[ {\begin{array}{*{20}c}
   {e^{ - i\left( {u_{1k} l + v_{1k} m} \right)} }  \\
    \vdots   \\
   {e^{ - i\left( {u_{N_{\rm tel} k} l + v_{N_{\rm tel} k} m} \right)} }  \\
\end{array}} \right]}  \\
~\\
	{{\bf{B}}_f  = \left( {\begin{array}{*{20}c}
   	{B_f \left( {{\bf{s}}_1} \right)} & {} & {}  \\
   	{} &  \ddots  & {}  \\
   	{} & {} & {B_f \left( {{\bf{s}}_d} \right)}  \\
	\end{array}} \right)}  \\

\end{array}
}
and where $k$ is the time--ordered visibility number, $\bf{s}$ is the source position vector on the celestial sphere 
and $f$ the observing frequency channel number.

\begin{figure*}

\includegraphics[width=0.9\textwidth]{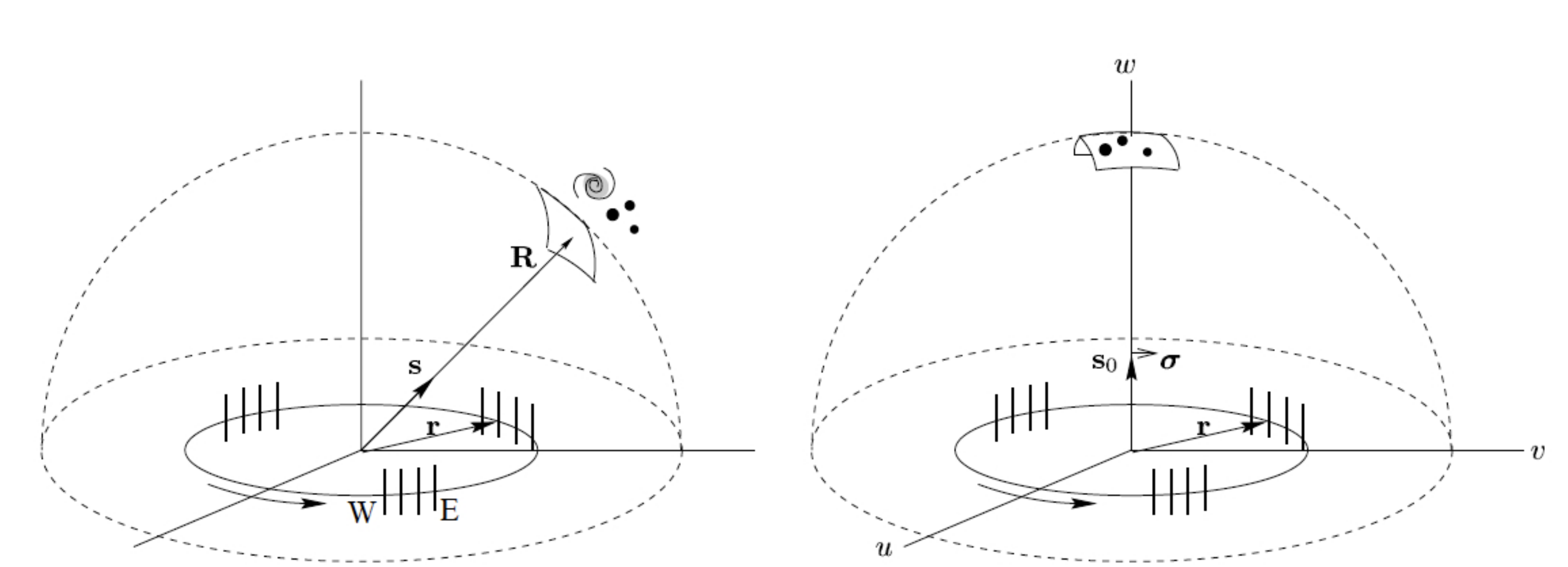}

\caption{The sky brightness distribution before and after geometrical delay compensation, as seen by an 
interferometer. $\bf{R}$ is the source position vector and $\bf{s}$ the relevant unit vector in the direction of $\bf{R}$. $\bf{r}$ is the baseline vector.}
\label{svector}
\end{figure*}

The vector functions ${\bf{a}}_{k,f}$ are called array response vectors in array signal processing and they are 
frequency dependent, but also time dependent in this case due to the rotation of the Earth \citep{vdv00b,vtrees}. They describe 
the response of an interferometer to a source at direction ${\bf s}=(l,m)$ (see Figure \ref{svector}).

The above formalism is trivial as long as the positions of the telescopes are well known. In reality though, the 
response of the array is not perfect: telescopes are not omni-directional antennas, but each one has its own 
properties (i.e. complex beam-shape and gain etc.). In this case the array response vectors must be redefined as 
\begin{equation}
	{{\bf{a}}_{k,f} ({\bf{s}}_i ) =
	{\left( {\begin{array}{*{20}c}
   	{{\cal{A}}_1 \left( {{\bf s}_{i}} \right)}   \\
   	 \vdots   \\
   	{{\cal{A}}_{N_{tel} } \left( {{\bf s}_{i}} \right)}  
	\end{array}} \right)} \odot
	\left[ {\begin{array}{*{20}c}
		
   	{e^{ - i\left( {\bf u}_{1k} \cdot {\bf s}_{i} \right)} }  \\
    	\vdots   \\
   	{e^{ - i\left( {{\bf u}_{N_{\rm tel} k} \cdot {\bf s}_{i} } \right)}} 
	\end{array}}\right]},
\end{equation}
where $\odot$ indicates a Hadamard (element-wise) product.
The source structure can also vary with frequency. Finally, most of the received signal consists of additive noise. 
When the noise has zero mean and is independent among the antennas (spatially white), then 
\[
{\bf{R}}_{k,f}  = {\bf{A}}_{k,f} {\bf{BA}}_{k,f}^\dag   + \sigma_{f} ^2 {\bf{I}}
\]
 Noise can be assumed to be Gaussian 
in radio-interferometers, like LOFAR. We will assume this in the remainder of this paper, and might address
non-Gaussian \citep{tms} time-varying noise in future publications. Actually, system noise 
is slightly different at each receiver. It is reasonable to assume that noise is spatially white: the noise covariance 
matrix is diagonal. 

\subsection{The Polarization Measurement Equation}

The equations above describe the relation of the visibilities to the total intensity of the source, i.e. the classical 
Stokes $I$. To take polarization into account, several modifications are required. The use of the polarization 
information is of great importance  for several reasons. First, this information provides insight in the physical 
processes that might exist in the astronomical object of interest.  Second, modern 
telescopes like LOFAR are also inherently polarized. Third, using more information  enhances the 
result of the data reduction process. The study of polarized light is therefore becoming an increasingly important 
issue in astrophysics \citep{tinbergen}. 

Many matricial models have been developed to study the polarization properties of light. A proper description of 
the polarization properties of light relies on the concept of the coherency matrix \citep{bornwolf,hbs1}.  This mathematical 
formalism holds for every band of the electromagnetic spectrum. A usual assumption is monochromatic light, but 
polychromatic light behaves as monochromatic for time intervals longer than the natural period and shorter than 
the coherence time \citep{gil07,barakat81}.
Our mathematical model will be based on those introduced in radio-astronomy by \citet{hbs1}.

\subsubsection{The Electric-Field Vector and Coherency Matrix}

The effects of linear passive media on the propagated photons can be represented by linear transformations of 
the electric field variables. The nature of those effects, the spectral profile of the light and the chromatic and 
polarizing properties of the medium through which light passes, all affect the degree of mutual coherence. In 
general coherent interactions can be represented by the Jones calculus \citep{jones41,jones42,jones48}, while 
incoherent interactions of polychromatic light require the Mueller calculus \citep{barakat81}, since the loss of 
coherence needs more parameters to be described. The two  components of the electric field (e.g. those 
received at two dipoles) can be arranged as the components a 2$\times$1 complex vector:
\e{
{\bf{e}}\left( t \right) = \left( {\begin{array}{*{20}c}
   {E_x \left( t \right)}  \\
   {E_y \left( t \right)e^{i\delta \left( t \right)} }  \\
\end{array}} \right)
}
where $\delta \left( t \right)$ is the relative phase. This vector includes all information about the temporal 
evolution of the electric field. When the parameters have no time dependence this is called the {\sl Jones vector}. 
Moreover, the coherency (or polarization or density) matrix of a light beam contains all the information about its 
polarization state. This Hermitian $2 \times 2$ matrix is defined as
\begin{equation}
\begin{split}
{\bf{C}} \equiv \left\langle {{\bf{e}}\left( t \right) \otimes {\bf{e}}^\dag  \left( t \right)} \right\rangle  =\\
\left( {\begin{array}{*{20}c}
   {\left\langle {e_1 \left( t \right)e_1^* \left( t \right)} \right\rangle } & {\left\langle {e_1 \left( t \right)e_2^* \left( t 
\right)} \right\rangle }  \\
   {\left\langle {e_2 \left( t \right)e_1^* \left( t \right)} \right\rangle } & {\left\langle {e_2 \left( t \right)e_2^* \left( t 
\right)} \right\rangle }  \\
\end{array}} \right)
\end{split}
\end{equation}
This is the coherency matrix of the perpendicular dipoles of a single LOFAR HBA antenna. $ \otimes $ stands for 
the Kronecker product and the brackets indicate averaging over time \citep{boonstra05}. The coherency matrix is a 
correlation matrix whose elements are the second moments of the signal. Using the ergodic hypothesis the 
brackets can be considered as ensemble averaging. Due to its statistical nature its eigenvalues ought to be non-negative.  The normalized version of this matrix ${\bf{\hat C}} = \frac{{\bf{C}}}{{{\rm tr}({\bf{C})}}}$
contains information about the population and coherencies of the polarization states \citep{fano57}. This object is the 
equivalent of the single brightness point in the scalar version of the theory.

\subsubsection{The Stokes Parameters}

Above we discussed the statistical interpretation of the coherency matrix. Now we can also introduce a 
geometrical description: the measurable quantities Stokes $I$, $Q$, $U$ and $V$ arise as the coefficients of the 
projection of the coherency matrix onto a set of Hermitian trace-orthogonal matrices, the generators of the unitary 
SU(2) group plus the identity matrix (see Appendix A1). Parameters with direct physical meaning can be derived from the 
corresponding measurable quantities. The Stokes parameters are usually arranged as a $4 \times 1$ vector,  
$${\bf s} =\left(\begin{array}{c} I\\ Q \\ U \\V \end{array}\right).$$ An alternative notation, that will be derived in 
Appendix A is the $2 \times 2$ Stokes matrix: 
\begin{equation}
	{\bf S} = \frac{1}{2}\left(\begin{array}{*{20}cc}
	I + Q & U -i V \\
	U + i V & I-Q
		\end{array}\right) \equiv {\bf C},
\end{equation}
which relates the measured coherency matrix quantities to the Stokes parameters \citep{bornwolf}.

\subsubsection{The Jones Formalism}

An adequate method to describe a non-depolarizing system is the Jones formalism. It represents the effects on 
the polarization properties of an EM wave after the interaction with such a system. For passive, pure systems, 
the electric field components of the light interacting with them is given by the corresponding Jones matrix $\bf J$, 
$${\mathbf{e}}' = {\mathbf{Je}}.$$ As both the initial and final fields can fluctuate, it is useful to describe the 
properties of partially polarized light with the coherency matrix. Thus, 
\e{\begin{aligned}
  {\mathbf C}' = \left\langle {{\mathbf{e}}' \otimes {\mathbf{e}}'^\dag  } \right\rangle  = \left\langle {({\mathbf{Je})} 
\otimes  {({\mathbf{Je}})} ^\dag  } \right\rangle  \\ 
   = \left\langle {{\mathbf{Je}} \otimes {\mathbf{e}}^\dag  {\mathbf{J}}^\dag  } \right\rangle  = {\mathbf{J}}\left\langle 
{{\mathbf{e}} \otimes {\mathbf{e}}^\dag  } \right\rangle {\mathbf{J}}^\dag   = {\mathbf{JCJ}}^\dag   \\ 
\end{aligned} 
}
As we are dealing with interferometry, the two $\bf{J}$ matrices can come from two different telescopes.
The effects on the electric field vectors in the coherency matrix ${\bf C}'$ can be written as an operation of
these Jones matrices on the original unaffected coherency matrix ${\bf C}$. 
As we already mentioned, the coherency matrix can also be written as a four vector with ${\mathbf{c}} = 
\left( {\left\langle {e_1 e_1^* } \right\rangle ,\left\langle {e_1 e_2^* } \right\rangle ,\left\langle {e_2 e_1^* } \right
\rangle ,\left\langle {e_2 e_2^* } \right\rangle } \right)$. This vector is related to the Stokes vector via \citep{parke}
\e{{\mathbf{s}} = {\mathbf{Lc}} {\text{~~~with~~~}}{\mathbf{L}} = \left( {\begin{array}{*{20}c}
   1 & 0 & 0 & 1  \\
   1 & 0 & 0 & { - 1}  \\
   0 & 1 & 1 & 0  \\
   0 & -i & { i} & 0  \\
 \end{array} } \right).}
The matrix has the following property ${\mathbf{L}}^{ - 1}  = {\raise0.7ex\hbox{$1$} \!\mathord{\left/
{\vphantom {1 2}}\right.\kern-\nulldelimiterspace}
\!\lower0.7ex\hbox{$2$}}{\mathbf{L}}^ {\dag}  $. 
Using the properties of the Kronecker product we then find, in terms of Stokes parameters, that
\e{\begin{aligned}
  {\mathbf{s}}' = {\mathbf{L}}\left\langle {{\mathbf{Je}} \otimes \left( {{\mathbf{Je}}} \right)^\dag  } \right\rangle  = 
{\mathbf{Ns}} \\ 
  {\mathbf{N}} = {\mathbf{L}}\left( {{\mathbf{J}} \otimes {\mathbf{J}} } \right){\mathbf{L}}^{ - 1}  \\ 
  \text{with} \\ 
  {\rm \mathbf{N}}_{kl}  = \frac{1}
{2}{\rm tr}\left( {{\mathbf{\sigma }}_k {\mathbf{J\sigma }}_l {\mathbf{J}}^ {\dag}  } \right). \\ 
\end{aligned} }
where $\mathbf{\sigma}_i$ are the Pauli matrices (Appendix A1). A Jones matrix can represent a physically realizable state as long as the transmittance condition (gain or 
intensity transmittance) holds; that is, the ratio of the initial and final intensities must be $0 \leqslant g \leqslant 1$. 
The reciprocity condition describes the effect when the output signal follows the path in the inverse order. For 
every proper Jones matrix ${\mathbf{e}}' = {\mathbf{J}}^ {\dag}  {\mathbf{e}}$. This result does not hold when 
magneto-optic effects are present. In this case the Mueller--Jones matrices have to be used.  If a Jones matrix 
represents a physically realizable state the reciprocal matrix also represents a physical effect.  
%


\section{The LOFAR measurement equation}

LOFAR stations consist of tiles of 4$\times$4 crossed dipoles which allow for full Stokes measurements. If we 
have $N_{\mathrm{tel}}$ stations each with two polarization degrees of freedom, then the $2 N_{\mathrm{tel}}$ electric fields can be 
stacked into a single vector and the correlation matrix can be generalized to the following form
\begin{eqnarray}
{\bf{V}}  & = & 
	\left( {\begin{array}{*{20}c}
   	{\left\langle {E_x \left( {{\bf{r}}_i } \right)E_x^* \left( {{\bf{r}}_j } \right)} \right\rangle } & {\left\langle {E_x 
\left( {{\bf{r}}_i } 	\right)E_y^* \left( {{\bf{r}}_j } \right)} \right\rangle }  \\
   	{\left\langle {E_y \left( {{\bf{r}}_i } \right)E_x^* \left( {{\bf{r}}_j } \right)} \right\rangle } & {\left\langle {E_y 
\left( {{\bf{r}}_i } 	\right)E_y^* \left( {{\bf{r}}_j } \right)} \right\rangle }  \\
	\end{array}} \right) \nonumber \\
	& =  &  \left( {\begin{array}{*{20}c}
   {{V}_{xx} } & {{V}_{xy} }  \\
   {{V}_{xy} } & {{V}_{yy} }  \\
	\end{array}} \right)
\end{eqnarray}
for a linear polarization basis. Note that each element of $\bf{V}$, ${V}_{ij}$ is not Hermitian for $i \ne j$, 
but ${{V}}_{ij}  = {{V}}_{ji}^\dag  $, so that $\bf{V}$ remains Hermitian \citep{hbs1}.  Here we have denoted explicitly the 
correlation from the X and Y oriented dipoles.

Let ${\bf{A}}_{il}$ be the position dependent polarization multiplication matrix. This is the array response matrix (as 
it is termed in the language of communication theory) or in the Hamaker formalism the Jones matrix. The array response vector must take into account the different 
physical and instrumental effects that affect the signal through its path from the source to the recorder, like 
ionospheric Faraday rotation, parallactic offsets, the geometric delay and instrumental gains (and leakages). 
These effects are represented by the Jones matrices. 
So we can write the measurement equation in the form \citep{boonstra05,leshem04}
\[
{\bf{V}}_{ijl}  = {\bf{A}}_{il} {\bf{B}}_l {\bf{A}}_{jl}^\dag  + \textbf{N}_{\rm noise} 
\]
 where the different effects $\bf{A}$ in the array response vector must be introduced in the exact order in which they affect the signal.  The index $l$ is the pixel number and the index $i,j$ represent antenas $i$ and $j$ respectively
The algebra of complex Jones matrices is obviously non-commutative, i.e.\ the ordering of the matrices matters. The 
physical meaning of this is that the results of the different effects on the incoming electromagnetic wave are not 
linear. 

\subsection{Individual Jones matrices}

In this section we give a brief overview of the instrumental parameters (Jones matrices) which are used for our
data simulations, their importance, and which parameters we use and perturb.

\paragraph*{\textbf{F}: Ionospheric Faraday Rotation}

The ionosphere is birefringent, such that one handedness of circular polarization is delayed with respect to 
the other, introducing a dispersive phase shift in radians
\[
\Delta \phi  \approx 2.62\times10^{-13}{\rm{  }}\lambda ^{\rm{2}} \int {B_{||} n_e \mathrm{d}s} {\rm  (SI)}
\]
It rotates the linear polarization position angle and is more important at longer wavelengths, at times close to the solar 
maximum and at Sunrise or Sunset, when the ionosphere is most active and variable \citep{bo08a,bo08b,bo07a,bo07b,spoel1,spoel2,spoel3,spoel4,spoel5}.
The Faraday rotation Jones matrix is a real-valued rotation matrix:
\[
\bf{F}  = \left( {\begin{array}{*{20}c}
   {\cos \Delta \phi } & { - \sin \Delta \phi }  \\
   {\sin \Delta \phi } & {\cos \Delta \phi }  \\
\end{array}} \right)
\]
To model the Total Electron Content (TEC) \citep{vdtol}, we use a Gaussian random field with Kolmogorov-like 
turbulence. We use a cut-off at scales that correspond to the maximum and minimum baselines of 
WSRT (Westerbork Synthesis Radio Telescope), similar to those
of LOFAR. Initial model parameters have been obtained from WSRT data, taken at similar frequencies, 
baselines and time intervals to our planned LOFAR observations. Moreover, the WSRT is located approximately 50 
km from the LOFAR core in Exloo and as such the WSRT provides a good test-bed for forthcoming LOFAR 
observations and expected
ionospheric effects. 

\paragraph*{\textbf{P}: Parallactic Angle}

This matrix describes the orientation of the sky in the telescope 's field of view, or similarly the projection of the dipoles onto
the sky. It has the mathematical structure of a rotation matrix and rotates the position angle of linearly polarized 
radiation
incident on the dipoles.  It should in principle be analytically and deterministically known, and its variation 
provides leverage for determining polarization-dependent effects. It can also be used at the initial stages to 
determine dipole orientation errors.
\[
{\rm{ }}{\bf{P}} = \left( {\begin{array}{*{20}c}
   {\cos \chi } & { - \sin \chi }  \\
   {\sin \chi } & {\cos \chi }  \\
\end{array}} \right)
\]
where $\chi$ is the parallactic angle.
We shall include this matrix in the antenna voltage pattern matrix.

\paragraph*{\textbf{E}:  Antenna Voltage Patterns}

\begin{figure}
	\centering
\caption{The design of a LOFAR HBA dipole antenna element.}
		\includegraphics[width=0.5\textwidth]{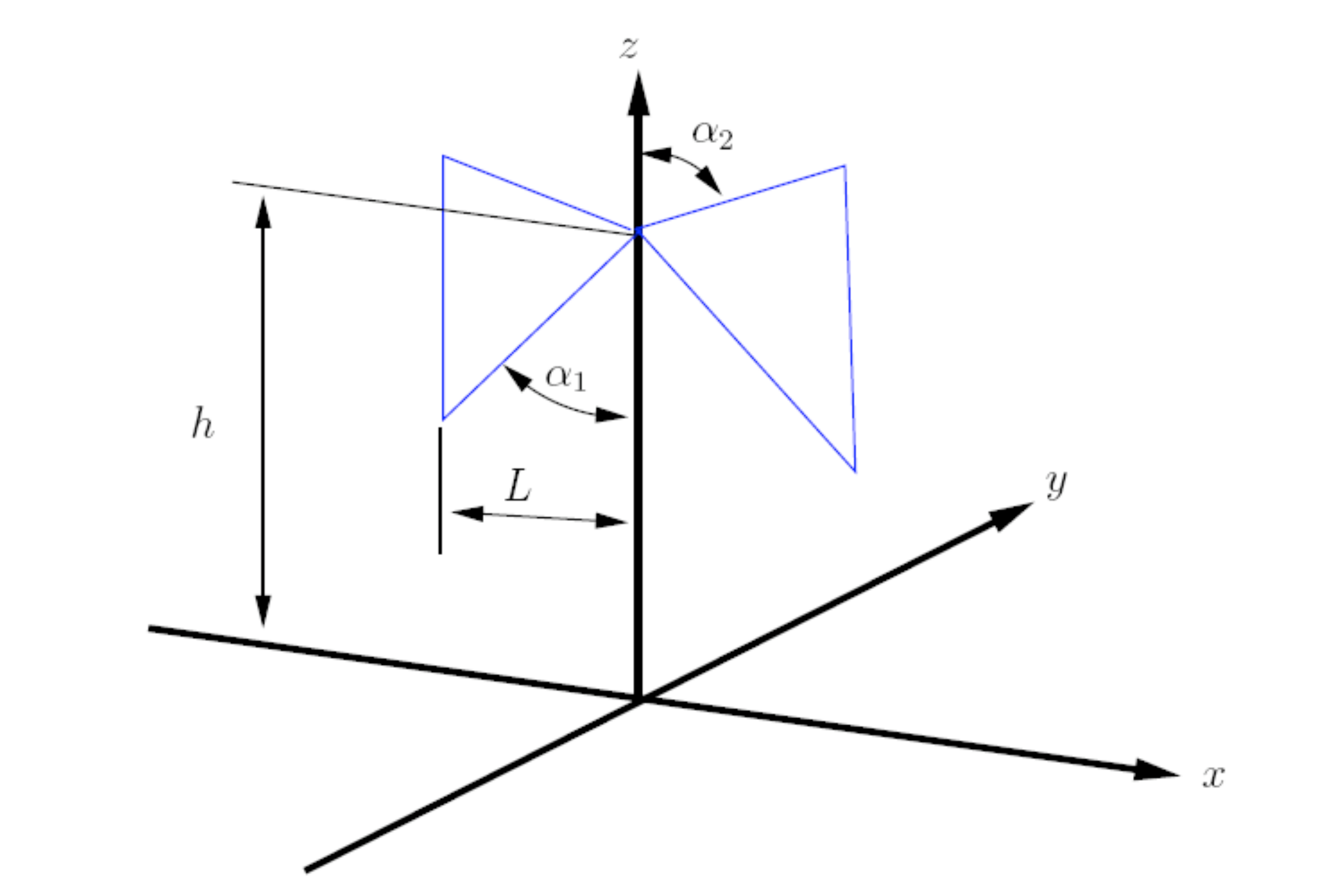}
	
	\label{fig:hba}
\end{figure}

The LOFAR telescope HBA stations consists of crossed bowtie dipole pairs. Like every antenna, they have a 
directionally dependent gain which is important when the angular size of the region on the sky is comparable to $ \sim \lambda /D$, where $D$ is the station diameter. 
At low frequencies when the radio sky is dominated by  point sources, wide-field techniques are required as well. 
To determine 
the antenna voltage pattern to first order, we assume an analytic model of the dipole pairs: the basic parameters 
as shown in Fig.~\ref{fig:hba} have the following values: \rm{L}=0.366 \rm{m}, \rm{h}=0.45 \rm{m}, $\alpha_{1}$= 
50 \rm{deg} and $\alpha_{2}$ = 80 \rm{deg}.

As the Earth rotates, the dipoles rotate with respect to the sky. This causes the polarization coordinate system to 
rotate (unlike e.g. the WSRT which has an equatorial mount) and we must thus correct for this effect. For a pair 
of crossed 
dipoles along the X and Y axes the relevant Jones matrix \textbf{E} can be written as:
\[
{\bf{E}} = \left( {\begin{array}{*{20}c}
   {E_\theta  \left( {\frac{\pi }{2} - \theta ,\phi  - \frac{\pi }{4}} \right)} & {E_\phi  \left( {\frac{\pi }{2} - \theta ,\phi  - 
\frac{\pi }{4}} \right)}  \\
   {E_\theta  \left( {\frac{\pi }{2} - \theta ,\phi  - \frac{{3\pi }}{4}} \right)} & {E_\phi  \left( {\frac{\pi }{2} - \theta ,\phi  - 
\frac{{3\pi }}{4}} \right)}  \\
\end{array}} \right)
\]
where $\phi$ and $\theta$ are the polar and azimuthal angles of the beam pattern.
We assume that the X and Y dipoles have the same beam pattern.  Because of the spatial distribution of the 
dipoles within a station, each dipole has a different delay for the reception of an electromagnetic wave coming 
from a source. If we assume a narrowband system, one can correct for this by shifting the phase for the signal of 
each dipole element of the stations
(i.e.\ introducing an effective delay). We have an analytical model for the HBA antenna element beam and we assume that the dipoles are uniformely distributed to form a circular array. Thus, we ignore the tile structure, for the purposes of the current paper. If we need to form a station-beam in an arbitrary direction inside the dipole/
tile beam, we can do it by properly weighting the signals of each element  and adding delays. Another 
simpler way to proceed, which is sufficient for the purposes of this paper, is to calculate the beam shape of an 
axially symmetric, uniformly distributed array of elements and then multiply it with the primary element pattern 
that we calculated above. We also assume that the thin-wire approximation holds.

\paragraph*{\textbf{D}: Polarization Leakage and Instrumental Polarization}

Radio--interferometers usually measure the full set of Stokes parameters simultaneously. Measuring the 
polarization properties of Galactic diffuse emission as well as the extragalactic sources is an important aspect of 
the EoR-KSP. The polarization fraction of most astronomical sources is typically low, of the order of few per cent. 
Measuring it with accuracy is thus challenging, but important in order to extract scientific information from it. 
Cross-polarization or polarization leakage is an instrumental contamination that couples orthogonal Jones 
vectors. The dipoles are not ideal, so orthogonal polarizations are not perfectly seperated. A slight tilt between 
the dipoles, for example, can change the polarization reception pattern. This must be taken into account, 
especially when we aim to reach  a high dynamic range in the images. It is supposed to be of the order of less 
than a few per cent and because it is a geometrical factor, we expect that it is frequency dependent. The relevant 
Jones matrix can be written in the form of a unitary matrix \citep{heiles,heiles1,sanjay01c,reid}:

\e[DJones]{{\bf{D}} = \left( {\begin{array}{*{20}c}
   {\sqrt {1 -  d}} & { - \sqrt d \, e^{i\psi } }  \\
   {\sqrt d \, e^{-i\psi } } & {\sqrt {1 -  d}}  \\
\end{array}} \right)}
where $d$ is a generally small constant of the order of $10^{-6}$ \. This is a first-order approximation to 
behaviour that might prove more complex. However, it is very important to know this parameter as it can convert 
unpolarized radiation, such as the EoR signal, into polarized radiation. This matrix has almost the structure of 
a rotation matrix. The difference is the inclusion of the phase $\psi$. When $\psi$ equals zero,  polarization 
leakage converts $Q$ to $U$ or $E$ to $B$-modes [rotation across the equator of the Poincar\'e sphere \citep{heiles}]. When the 
phase term is significant, polarization leakage leads to mixing $Q$, $U$ and $V$. Polarization leakage manifests itself as 
closure errors in parallel hand visibilities and the leakage-induced closure phase and the Pancharatnam phase of 
optics \citep{berry}. This phase arises when the state of polarization of the light is transformed following a closed path in the 
space of states of polarization, which is known as the Poincar\'e sphere \citep{bornwolf}. Another type of distortion is the 
instrumental polarization. The dipole and station designs as well as the processing done can contribute to such 
effects. LOFAR uses linearly polarized dipoles. The interferometric measurement of the Stokes parameters U 
and V using two different dipoles is made by forming the ``cross-handed'' products of the signals. If for any 
reason the signals are received with different gains, this would lead to a certain fraction of polarization. This can 
be modelled as: 
\e{{\bf{D}}_\mathrm{p}  = \left( {\begin{array}{*{20}c}
   {d_1 } & 0  \\
   0 & {d_2 e^{i\psi } }  \\
\end{array}} \right),}
 where $d_1$, $d_2$ are the different X and Y gains. We assume that the instrumental polarization axis lies 
along the axis defined by the matrix basis to write this equation. In such case $Q$ does not leak into $U$, but still $I$ 
leaks onto $Q$. With a carefully chosen basis (see Appendix A) $I$ does not leak into $U$. 
\paragraph*{\textbf{G}: Complex baseline-based electronic gain}

This matrix accounts for all amplitude-phase- and frequency-independent effects of the station electronics. It 
shows a slow variation of the order of 1--2\% over a day. It has the form:
\[
{\bf{G}} = \left( {\begin{array}{*{20}c}
   {g_X } & 0  \\
   0 & {g_Y }  \\
\end{array}} \right),
\]
where $g_X$ and $g_Y$ are complex numbers. It is one of the most common calibration parameters and the one
commonly solved for in the classical self-calibration loop.
 
\paragraph*{\textbf{B}: Bandpass}

Compensating for change of gain with frequency is called bandpass calibration. This matrix is similar to \textbf{G} 
and describes the frequency-dependence of the antenna electronics. Station digital  poly-filters, used to select 
the frequency passband, are not perfectly square, but they are deterministically known and their bandpass 
behavior is expected to be quite stable over large periods of time. Its generic form is: 
\[
{\bf{B}} = \left( {\begin{array}{*{20}c}
   {b_x (f )} & 0  \\
   0 & {b_y (f)}  \\
\end{array}} \right)
\]
The $b$ are real numbers. We model both {\bf B} and {\bf G} parameters as slowly time-varying functions with additive white noise as the initial comissioning tests for LOFAR suggest.

\paragraph*{\textbf{K}: Fourier kernel}

This term has traditionally been called the Fourier kernel. For larger fields of view and/or longer baselines the 2--D Fourier transform relationship between the sky and the measured visibilities is no longer a good 
approximation. This term describes a phase shift that accounts for the geometrical delay for the signals received 
by telescopes $i$ and $j$. Let $x, y$ and $z$ be the antenna positions in a coordinate system that points to the 
West, South and the Zenith respectively. We can provide the directions on the sky in terms of the directional cosines 
$l, m, n$. We assume that the sky is a unit sphere so that $l^2  + m^2  + n^2  = 1$ \citep{tms}.  Since the X and Y dipoles are co-located, the relevant Jones matrix takes 
the following simple form:
\[
{\bf{K}} = e^{{-i\left[ {ul + vm + w\left( {\sqrt {1 - l^2  - m^2 }  - 1} \right)} \right]} }\left( {\begin{array}{*{20}c}
   1 & 0  \\
   0 & 1  \\
\end{array}} \right)
\]
This matrix should also  be well determined, but solving for it can help to estimate the antenna positions, the 
electronic path lengths, the clock errors etc. For this the concept of the array manifold is useful \citep{manikas}
It also provides astrometric information by solving for $l$, $m$ and $n$, which 
are the direction cosines of the sources on the sky.

\subsection{The Final Form of the Measurement Equation}

All the above effects can be combined in order to form the Hamaker--Bregman--Sault  
measurement equation. The equation looks like:

\begin{equation}
\begin{array}{l}
 V^{{\rm obs}}_{ij}  = \int\limits_{l,m,n} {\mathrm{d}l \mathrm{d}m\left( {{\bf{A}}_{i}  \otimes {\bf{A}}_{j}^{\bf{\dag}} } \right){\bf{c}}
\left( {l,m} \right)} , {\rm ~~~with} \\ ~ \\
 {\bf{A}}_{i}  = {\bf{K}}_{i} {\bf{B}}_{i} {\bf{G}}_{i} {\bf{D}}_{i} {\bf{E}}_{i} {\bf{P}}_{i} 
{\bf{T}}_{i} {\bf{F}}_{i}  \\~\\ 
 {\bf{A}}_{j}^{\dag} {\bf{ = F}}_{j}^{\bf{\dag }} {\bf{T}}_{j}^{\bf{\dag }} {\bf{P}}_{j}^{\bf{\dag }} 
{\bf{E}}_{j}^{\bf{\dag }} {\bf{D}}_{j}^{\bf{\dag }} {\bf{G}}_{j}^{\bf{\dag }} {\bf{B}}_{j}^{\bf{\dag }} 
{\bf{K}}_{j}^{\bf{\dag }}  \\ 
 \end{array}
\end{equation}
~\\
where  $\bf{c} = {\rm vec}(\bf C)$ is the 4$\times$1 source 
coherency vector in the $l, n, m$ system\footnote{Here we use the Kronecker product identity ${\bf{AXB = C}} 
\Leftrightarrow \left( {{\bf{B}}^T  \otimes {\bf{A}}} \right){\rm vec}\left( {\bf{X}} \right) = {\rm vec}\left( {\bf{C}} \right)
$. We can then solve for $\bf{X}$, when $\left( {{\bf{B}}^T  \otimes {\bf{A}}} \right)$ is invertible \citep{golub}. This way the 
measurent equation can be transformed into a linear equation.}.

If we assume that the sky can be described as a closely-packed collection of point-sources, then the 
measurement 
equation can be rewritten as:
\[
V^{{\rm obs}}_{ij}  = \sum\limits_{l,m} {{\bf{A}}_{i} \left( {l,m} \right){{\bf C}\left( {l,m} \right){{\bf 
A}}_{j}^{\dag}\left( {l,m} \right) }} 
\]
where $\bf{C}$ is the coherency matrix of a single point-source at ($l,m$). For a single point-source with the rest of the sky blank this simplifies 
further 
to $$
V^{{\rm obs}}_{ij}  = {\bf{A}}_{i} \left( {l,m} \right){\bf{C}}\left( {l,m} \right){\bf{A}}_{j}^{\dag} \left( {l,m} 
\right).$$
We must note that this equation is linear over the coherency matrix \textbf{C}, but not over the Jones matrices 
\textbf{A}. We reiterate that the invidual Jones matrices that form the matrix products \textbf{A}, are not in 
general commutative. Their order follows the signal path. The form of this equation is quite complicated. Every element in the new matrix is a non-linear function of all the parameters that 
appear in the various Jones matrices.  

It has been proposed \citep{hbs5,ham1} that we can self-calibrate the generic \textbf{A} matrix, apply 
post-calibration constraints to ensure consistency of the astronomical absolute calibrations, and recover full 
polarization measurements of the sky. This is an interesting idea for low-frequency arrays, where isolated 
calibrators are unavailable (due to the fact that such arrays see the whole sky). This just emphasizes the fact that 
calibration and source structure are tied together - one cannot have one without the other.

\subsection{Additive errors}

The ultimate sensitivity of a receiving system is determined principally by the system noise. 
The discussion of the noise properties of a complex receiving system like LOFAR can be lengthy \citep{lopez,PP07}, so
we concentrate for our purposes on some basic principles. The
theoretical $rms$ (root mean square)  noise level in terms of flux density on the final  
image is given by \citep{tms}

\begin{equation}
\sigma _{\rm{noise}} = \frac{1}{{\eta _s }} \times \frac{{\rm{SEFD}}}{{\sqrt {N
\times (N - 1) \times \Delta \nu \times t_{{\mathop{\rm int}} } } }}  
\end{equation}
where $\eta _s$ is the system efficiency that accounts for
electronic, digital losses, $N$ is the number of substations, $\Delta
\nu$ is the frequency bandwidth, $t_{{\mathop{\rm int}} }$ is the  
total integration time and SEFD is the System Equivalent Flux Density. The system noise we assume to have two contributions: the
first comes from the sky and is frequency dependent ($\approx
\nu^{-2.55}$) \citep{shaver99,jelic08} and the second comes from from the receivers. The scaling of the $A_{eff}$, the effective collecting area of the antennae with frequency, introduces also a frequency dependence. We also assume that the 
distribution of noise over the uv--plane at one frequency is Gaussian, in both the real and imaginary part of the 
visibilities.  For
a 24-tile LOFAR core station, the SEFD will be around 2000~Jy ($\eta \sim 0.5$), depending on the final
design \citep{PP07}. This means that we can reach a sensitivity of 520  
mK at 150 MHz with 1 MHz bandwidth in one night  of 4 hours of
observation.  Accumulating data from a hundred nights of observations  
brings this number down to $\sim$52 mK. We will assume a constant noise
estimate of this magnitude for each pixel in the remainder of the paper. The
noise is uncorrelated between different telescopes and different
signals, so that we can write the noise matrix in the form:
\e{{\bf{N}} = \left( {\begin{array}{*{20}c}
   {\left\langle {{\bf{n}}_i {\bf{n}}_j^* } \right\rangle _1 } & {} & 0  \\
   {} &  \ddots  & {}  \\
   0 & {} & {\left\langle {{\bf{n}}_i {\bf{n}}_j^* } \right\rangle _{N_{\mathrm{tel}} } }  \\
\end{array}} \right)} where ${\bf{n}}_i  = \left\{ {n_i^{xx} ,n_i^{yy} } \right\}.$

We are now in a situation that we can start to simulate realistic data-sets for LOFAR.

\section{Realistic LOFAR-EoR-KSP Data Simulations}


In this section, we will describe the LOFAR instrumental response simulations and the inversion method used to 
enhance the result of the primary calibration. The simulation consists of two steps: (i) the forward step, where we use 
a realistic model of the LOFAR response function, including all instrumental effects and noise to generate simulated LOFAR EoR data, and (ii) an 
inversion step where we use the data model with realistic solutions and error range for the data model parameters, to obtain the 
underlying visibilities. In this subsection we shall consider the forward step. The data model is based on the 
Hamaker--Bregman--Sault measurement equation, as described in the previous sections,  and we use an 
``onion-layer'' approach to predict the instrumental response, as is suggested by the order of the Jones matrices 
in the  Measurement Equation. First, we must obtain a representation of the Fourier Transform of the sky, as it is 
perceived by an interferometer. To achieve this we  predict the values of the ${\bf{K}}$ Jones matrix. 
This includes calculating the $u$, $v$ and $w$ terms as well as the $l$, $m$ and $n$ direction cosines. For the 
purpose of this paper we calculated them  for six hours of integration and ten seconds of averaging. The centre 
of the sky maps is at $\delta_c= 52^\circ $ and $\alpha_c = 12$ hours.

The next step is to predict the `true' uncorrupted visibilities for different sky realizations. Those are the 
visibilities as seen by a perfectly calibrated and noiseless instrument. In this step, the source structure and the 
calibration problem are linked together in the visibilities \citep{selfcal1, hbs5,ham1,ham2,ham3}, thus we need to investigate the 
performance of the inversion method, using sky models with different complexities (see subsection 5.3). In this paper we 
apply the method to  the galactic diffuse emission 
superimposed on the cosmological signal. In the future we will also consider  a collection of point sources (this will be useful for the 
construction and improvement the Local Sky Model).

 The simulated maps between 200 and 110 ~\rm{MHz} were created 
at a high frequency resolution (100~\rm{kHz}) corresponding to the LOFAR EoR data-set resolution and were subsequently binned in frequency using a 1~{\MHz} moving  average box function. Each simulated visibility data-set has a size of approximately 100~\rm{MB}. We produced data-sets for 128 frequency channels (reduced from 320 that will be used during the real experiment).

The simulation, inversion and signal extraction algorithms are implemented in MATLAB, with extensive use of MATLAB executables files, to speed up the most often visited loops. Due to the parallel nature of the procedures, the Distributing Computing Toolbox is used, while many BLAS operations and the FFT are implemented using the relevant NVIDIA CUDA 2.0 libraries, and run on a single Tesla S870 system. For the rest we used a 16-way SMP workstation with 32 \rm{GB} of RAM and sufficient disk space.

\subsection{Cosmological signal and foregrounds simulations}

We start by discussing the  cosmological signal maps that we used in the simulation. The cosmological 21-cm 
signal is generated in a simplified manner, but sufficient for our purposes. We generate a contiguous cube of 
dark matter density from redshift $z\sim6$ to $z\sim12$ (corresponding to frequencies
of around 110 MHz to 200 MHz) as described in Thomas et al. (2008), from equally spaced outputs in time of an 
N-body simulation. We employ a
comoving 100 $h^{-1}$ ~Mpc , $256^3$ particles dark matter only simulation, using \textsc{GADGET 2}. We furthermore 
assume a flat $\Lambda$CDM universe and set the cosmological parameters to $\Omega_{\mathrm{m}} = 0.238$, $
\Omega_{\mathrm{b}} = 0.0418$, $\Omega_\Lambda = 0.762$, $\sigma_8 = 0.74$, $n_{\mathrm{s}}=0.951$ and
$h=0.73$, in agreement with the WMAP3 observations (\citealp{spergel07}).

The ratio between the baryons (basically atomic hydrogen and helium) to dark matter was set to $\Omega_{\mathrm{b}}/\Omega_{\mathrm{DM}} 
\approx 0.2$. Atomic hydrogen is assumed to follow an analytical ionization history of exponential nature as ${1+\exp(z-z_{\mathrm{ion}})}^{-1}$, where $z_{\mathrm{ion}}$ differs for each line-of-sight as $10 \pm N(0,1)$, where $N(0,1)$ is a 
normal distribution with mean zero and dispersion one. This is done to mimic a possible spread in redshift 
during which the reionization occurs. We plan to use a more realistic cosmological signal in the future, but for our 
current purposes (i.e. testing the inversion process), this is sufficiently complex.

For each of 
the foreground components, a
$5^\circ\times 5^\circ$ map is generated in the same frequency range (between 110 and 200~{\rm MHz}) 
pertaining to the LOFAR--EoR experiment. \emph{In this paper, we use only simulations of the Galactic diffuse 
synchrotron emission and of radio galaxies. More complex foreground realizations are under investigation and 
we will be considering them in future work.}

Of all components of the foregrounds, galactic diffuse synchrotron emission (GDSE) is by far the most  dominant 
and originates from the interaction between free electrons in the interstellar medium and the Galactic magnetic 
field. The intensity of the synchrotron emission can be expressed in terms
of the brightness temperature $T_{\mathrm{b}}$ and its spectrum is close to a featureless power law $T_{b}\sim 
\nu^{\beta}$, where $\beta$ is the
brightness temperature spectral index.

At high Galactic latitudes the minimum brightness temperature of the GDSE is about 20~K at 325~{\rm MHz} 
with variations of the order of 2
per cent on scales from 5--30~arcmin across the sky \citep[][]{debruyn98}. At the same Galactic latitudes, the 
temperature
spectral index $\beta$ of the GDSE is about $-2.55$ at 100~{\rm MHz} \citep{shaver99, rogers08} and steepens towards 
higher frequencies, but also gradually changes
with position on the sky \citep[e.g.][]{reich88, platania98}.

In the four dimensional (three spatial and one frequency) simulations of the GDSE, all its observed 
characteristics are included: spatial and
frequency variations of brightness temperature and spectral index, as well as brightness temperature variations 
along the line of sight. The
spatial distribution of the 3D amplitude and brightness temperature spectral index of the GDSE are generated as 
Gaussian random fields,
while along the frequency direction we assume a power law. The final map of the GDSE at each frequency is 
obtained by integrating the 4D cube along one spatial direction. All spatial and frequency properties of the 
simulated GDSE are normalized to match the values
of observations. In addition to the simulations of the total brightness temperature, polarized Galactic 
synchrotron emission
maps are also produced, that include multiple Faraday screens along the line of sight. For a detailed
explanation of the GDSE emission simulations in total and polarized brightness temperature see \citet{jelic08}.

\subsection{The uncorrupted visibilities}

The prediction of the visibilities for simple intensity distribution like point--sources and Gaussian distributions is 
straightforward. This is not the case for diffuse emission with complex structure.  There are two approaches in this case. In 
the discrete version of the problem, one can either use a very high resolution map and then assume that each 
pixel is a point source, taking into account the linearity of the ME over the brightness 
distribution, or one can find a proper representation of the sky distribution (i.e. shapelets) and use a carefully 
chosen gridding-convolution  function on the uv-plane. Image plane effects need to be computed locally, as they 
affect relatively small scales and thus, they have a small convolution footprint. This is the approach of the 
MeqTree/UVBrick software \citep{noordam06} which is used for predicting uv-data of extended sources or of multiple point sources. 
Despite the computational gains of this method, we shall use the more direct and precise multiple point-source 
prediction. This is done in order to preclude the introduction of spurious distortions on the simulated 
cosmological signal as well as to retain the high-resolution of the image-plane effect predictions. The sky 
brightness distribution can be split into patches in order to increase the speed of execution at the expense of 
additional phase--shifts (i.e. a shift in $(l,m,n)$, the phase center of the image, correponds to a phase shift of the Fourier kernel). In that case each discrete Fourier Transform and each frequency channel is independent, 
which yields an embarrassingly parallel computational process. 

In our matrix formulation the operation performed becomes 
$${\bf{V}}_{\rm true,f} \left( {u,v,w} \right) = \sum\limits_i^{N_{\rm pix} } 
{{\bf{K}}_{i,f} {\bf{B}}_{i,f} {\bf{K}}_{i,f}^\dag  },  $$ 
where ${\bf{B}}_{i,f}$ is coherency the matrix of 
pixel $i$ on a grid at each frequency $f$, and  ${\bf{K}}_{i,f}$ is the corresponding 
Fourier kernel in the direction of pixel $i$. The same approach holds for the point-sources as well,
which can have any given position. Hence, we can incorporate the different 
components of our sky model into a single formalism. Another reason for our preference for this method, is that the 
inversion of the above equation is equivalent to a deconvolution process. We shall discuss this in more 
detail in the following subsection.

\subsection{Simulations of the instrumental response}

In this section, we describe the different effects the instrument and ionosphere have on the observed
visibilities. These can be split in three categories: image-plane effects, uv-plane effects and noise. 

\subsubsection{Image--plane effects}

 As we saw, the array response matrix $
\bf{A}$ is composed of multiple effects, namely:

\begin{itemize}
 \item Faraday rotation
 \item Ionospheric phase fluctuations
 \item Antenna voltage pattern
 \item Polarization leakage and instrumental polarization
 \item Parallactic angle rotation
 \item Fourier kernel
\end{itemize}

The effects that depend on ${\mathbf s}= (l,m,n)$ are commonly called image-plane effects. We apply them directly on the 
images before taking into account the $\bf{K}$ Jones matrix. This is because they are direction-dependent gains 
or rotations of the coherency matrix. They transform each pixel on the maps in a different way. In the uv-plane, 
they enter as convolutions of the visibilities. For now, we ignore the parallactic angle $\bf{P}$ Jones matrix, 
because its terms should be known to very high precision a priori, and depend only on the rotation of the sky inside the 
station primary beam pattern. 

We only consider the beam pattern and the ionospheric Faraday rotation 
and phase delays. We use the LOFAR HBA dipole beam patterns, as well as the HBA station patterns provided 
in an ASTRON\footnote{http://www.astron.nl} technical report, by Yatawatta (2007). They depend on the element geometry and the station layout. Initial engineering results from Brentjens, Yatawatta and Wijnholds (unpublished) indicate that 
they are stable over a time interval of $\sim$8 hours.  We normalize the response at the 
phase center to one.  Slight misalignment of the dipoles, being not precisely orthogonal, would lead to a different 
polarized beam pattern response.  Because we are not correlating each dipole pair independently,  but we are 
actually doing aperture synthesis between entire stations, we do not have the information needed to correct this 
effect on a per dipole basis. In the simulations we assume that this effect is included in the errors of the relevant 
Jones matrices. We therefore include the net effect in the errors of the complex beam pattern. 

\begin{figure*}
\centering
\begin{tabular}{cc}
\epsfig{file=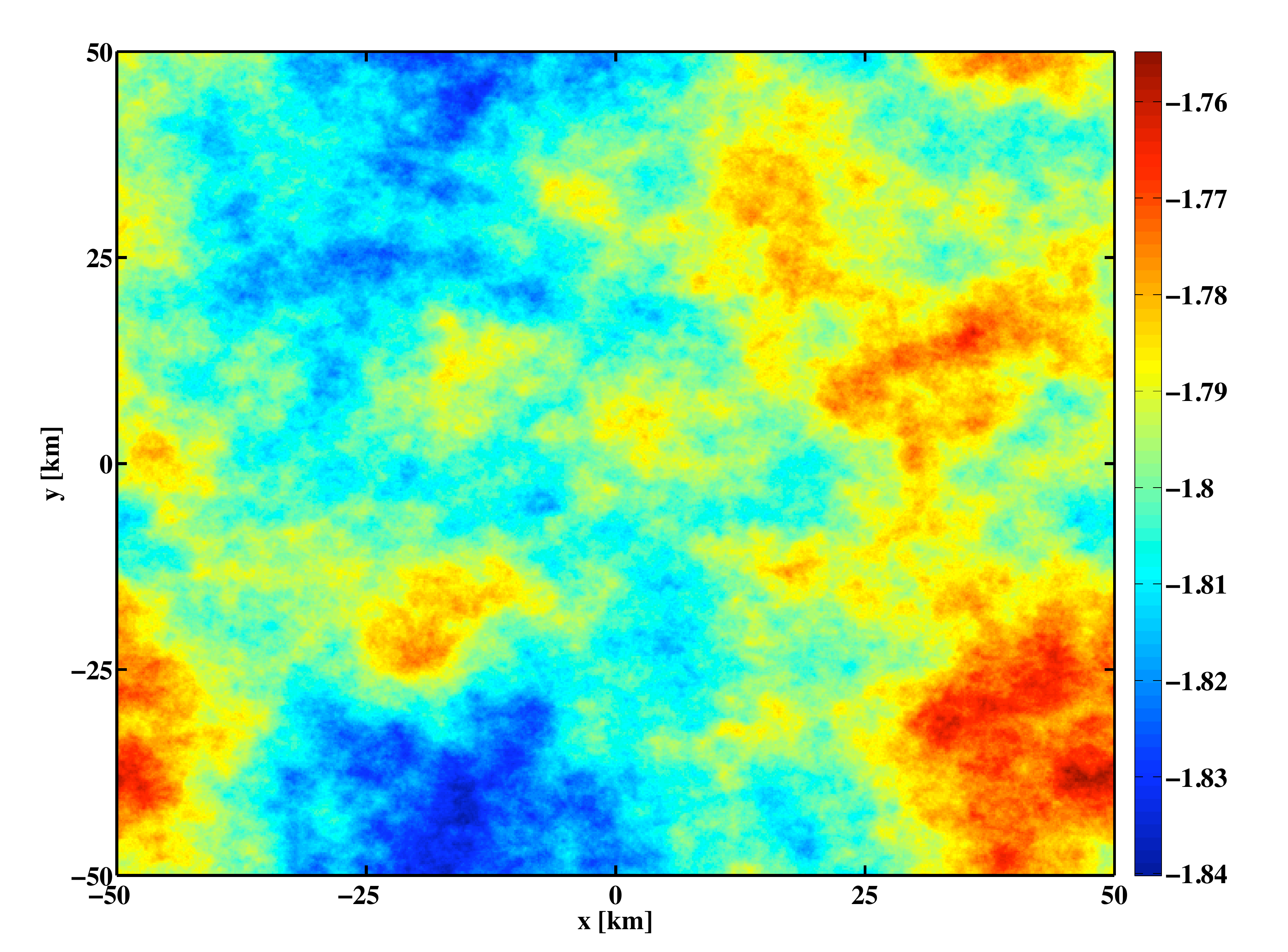,width=0.45\linewidth,clip=} &

\epsfig{file=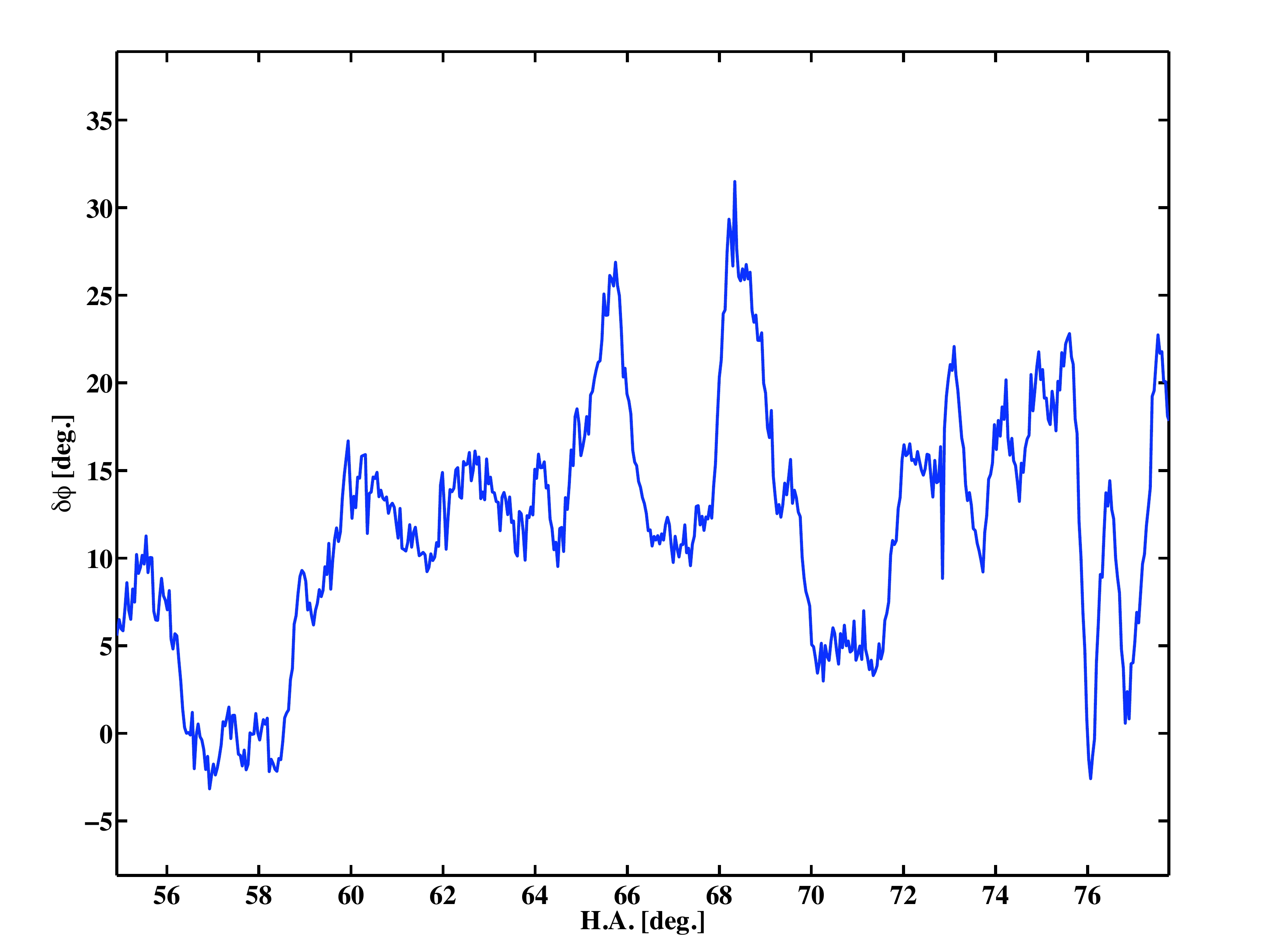,width=0.45\linewidth,clip=} 
\end{tabular}
\caption{ {\bf{Left:}} A simulated  map of the TEC of the ionosphere at a particulat time, above the LOFAR core. Each LOFAR station 
sees approximately twenty square kilometers of the ionosphere at an altitude of 250 \rm{km}. The values are 
displayed in \rm{TECU}s. {\bf{Right:}} Example of phase fluctuations caused by the ionosphere at a baseline of 100 metres as 
a function of Hour Angle}
\label{fig:ion}
\end{figure*}

The other two image-plane effects that we shall consider in this paper are the ionospheric phase and the 
Faraday rotation. We assume that the ionosphere consists of a single layer at an altitude of 250~\rm{km} above 
the ground. We assume a velocity with a mean of 200~$\mathrm{km~h^{-1}}$ and a variance of 10~$\mathrm{km~h^{-1}}$. This is 
done in order to simulate the turbulent temporary motion of the traveling ionospheric disturbances (TID). We use 
five directions with different velocities.  In such case the vertical and the slant Total Electron Contents (TEC) are 
equal. We assume that the TEC distribution is a random fields consistent with Kolmogorov-type turbulence. It has 
a mean value of 20 TECUs (1 TECU = $10^{16}$ electrons per $\mathrm{m}^2$) and a variance of $\sim$1 TECU as in Fig \ref{fig:ion}. Each 
station sees approximately 20$\times 20$ square kilometers of the ionosphere, depending on the frequency. We simulated a 
larger patch, though, of 50$\times 50$ square kilometers. The ionospheric phase-delay introduced would be
\e{\Delta \tau _{\mathrm{ion}}  =  - \frac{{40.3 {\rm m}^3 {\rm s}^{ - 2}  \cdot {\rm TEC_{src}} }}{{ \nu^2 }},} where ${\rm TEC_{src}}$ is the 
TEC along the line of sight, $c$ is the speed of light and $\nu$ is the observing frequency \citep{spoel4}. The form of 
the ionospheric delay Jones matrix is then $${\bf{Z}} = {\bf{I}}e^{i\Delta \tau _{\mathrm{ion}} } $$ and the Ionospheric Faraday rotation is

 \e{
{\bf{F}} = \left( {\begin{array}{*{20}c}
   {\cos \left( {{\raise0.7ex\hbox{${\rm RM}$} \!\mathord{\left/
 {\vphantom {{\rm RM} {\nu^2 }}}\right.\kern-\nulldelimiterspace}
\!\lower0.7ex\hbox{${\nu^2 }$}}} \right)} & { - \sin \left( {{\raise0.7ex\hbox{${\rm RM}$} \!\mathord{\left/
 {\vphantom {{\rm RM} {\nu^2 }}}\right.\kern-\nulldelimiterspace}
\!\lower0.7ex\hbox{${\nu^2 }$}}} \right)}  \\
   {\sin \left( {{\raise0.7ex\hbox{${\rm RM}$} \!\mathord{\left/
 {\vphantom {{\rm RM} {\nu^2 }}}\right.\kern-\nulldelimiterspace}
\!\lower0.7ex\hbox{${\nu^2 }$}}} \right)} & {\cos \left( {{\raise0.7ex\hbox{${\rm RM}$} \!\mathord{\left/
 {\vphantom {{\rm RM} {f^2 }}}\right.\kern-\nulldelimiterspace}
\!\lower0.7ex\hbox{${\nu^2 }$}}} \right)}  \\
\end{array}} \right)
}
where $\rm RM$ is the rotation measure \citep{rms05}. The rotation measure in SI units is defined as \e{\mathrm{RM}=\frac{e^3}{8\pi^2\epsilon_0m^2c^4} \int_0^d n_e B_{||}\,\mathrm{d}s\ }
where $\epsilon_0$ is the vacuum permittivity, with $B_{||}$ the magnetic field, $m$ the mass of the electron, $e$ the charge of the electron and $n_e$ the electron density.
From the simulated maps we construct the coherency matrix for every 
pixel of the maps. We then apply the image--plane effects on the maps and apply the $\bf{K}$ Jones matrix to 
bring them to the uv-plane. 

\subsubsection{uv--plane effects}

The next category of instrumental effects is the uv--plane effects:

\begin{itemize}
 \item Complex gains.
 \item Frequency bandpasses. 
\end{itemize}
The LOFAR gains and especially the 
bandpasses are expected to be stable over the period of one night with temporal variations of the order of two 
per cent. The bandpass response is well known and also well 
behaved. For this reason, the bandpass and complex gain effects together are handled in a single Jones matrix 
$\bf{G}$. Actually, the bandpasses are the frequency dependent parts of the instrumental complex gains. The  
frequency dependent gain can be approximated as:
\e{
\begin{array}{*{20}c}
   {{\bf{G}}\left( {f,t} \right) = \left( {\begin{array}{*{20}c}
   {G_1 } & 0  \\
   0 & {G_2 }  \\
\end{array}} \right)}  \\~\\
   {G_{\{ 1,2\} }  = g_{\{ 1,2\} } \left( {1 + 10^{ - 2} \sin (\omega _{\mathrm{cts}} t)} \right)\left[ {1 + \gamma \left( {\nu - \nu_0 } 
\right)} \right]},  \\
\end{array}
}
where  $g_{\{1,2\}}$ are the complex gain coefficients, $\omega _{\mathrm{cts}}$ is the cyclic frequency that corresponds 
to the correlation time scale of the gain solutions, 
$\gamma$ is constant with a small value and $\nu_0=150$ \MHz.  

\begin{figure}
\centering

\epsfig{file=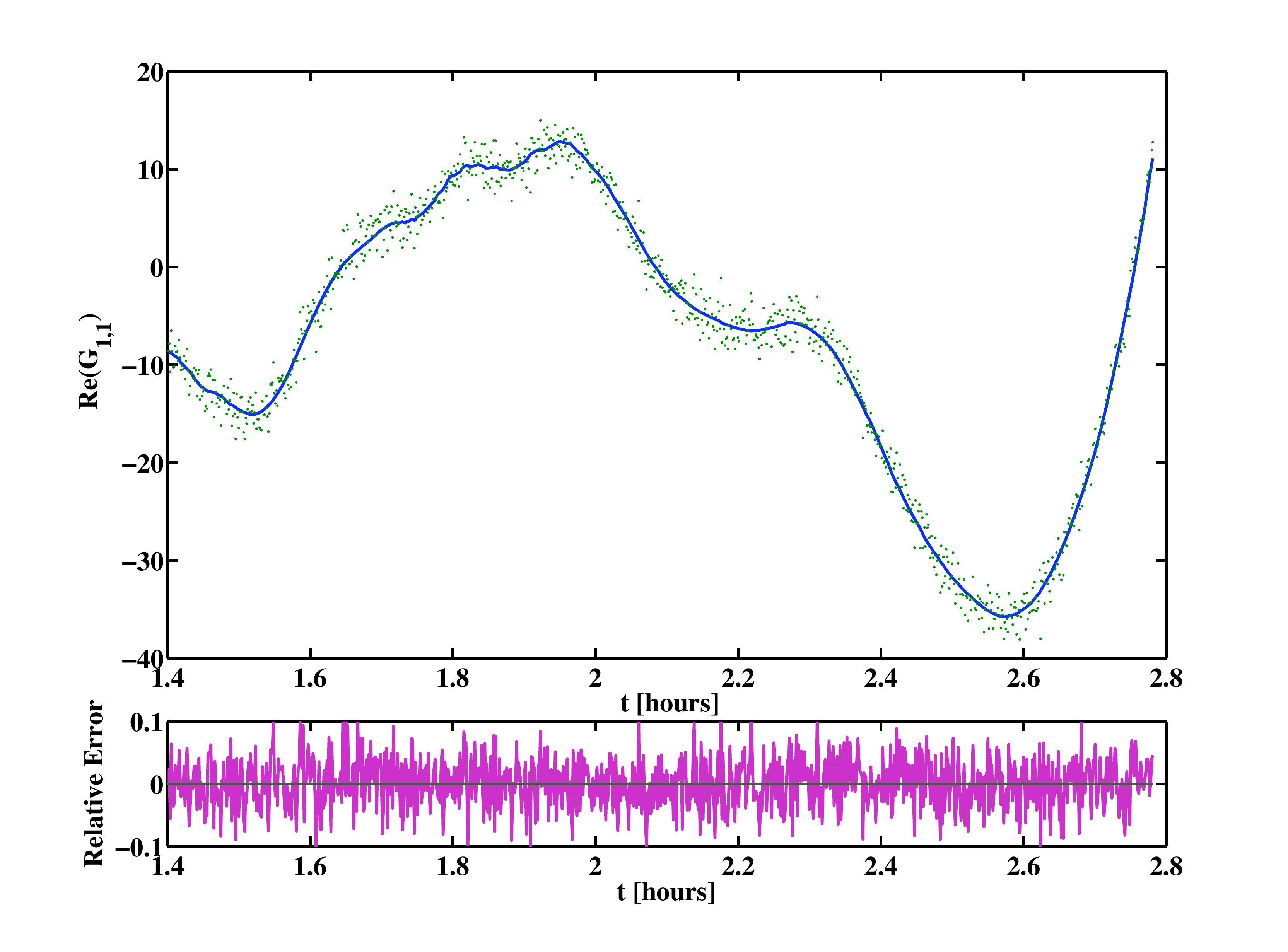,width=1.1\linewidth,clip=}

\caption{A plot of the real part of the complex gain for an XX correlation of a 500 \rm{m} baseline at $150~{\rm MHz}$ as function of time. The dots represent the recovered solution while the solid line the values used to corrupt the data. The relative error is plotted in the bottom panel.
}
\label{fig:gain}
\end{figure}

Since there are no HBA station 
cross-correlations available at the moment, we used the WSRT radio telescope solutions, scaled up as a function of 
collecting area.  For that we used the gain solutions from a deep survey of a few fields carried out by the LOFAR-EoR KSP team
during November 2007 (Bernardi et al., in preparation). The goals of this WSRT survey are to better understand  the Galactic foregrounds, at 
frequencies in a range similar to that spanned by the LOFAR high-band antennas, and to test the calibration 
pipeline. The survey was carried during 6$\times$12\rm{h} runs. One of the fields observed was the so-called Fan field 
(Bernardi, in preparation).
Finally, we model the polarization leakage following Equation \ref{DJones}, with $d=10^{-6}$ and $\psi=0$. In this 
paper we ignore the radio-frequency interference (RFI) contamination and we assume that the data have been 
sufficiently corrected. We will consider the implications of insufficient RFI flagging and its effects on the signal 
extraction  in the future.

Our final measurement equation, including all effects described above, can thus be written as:
\e[MEQ_PAPER]{
{\bf{A}}(\nu,t) = {\bf{G}}\left( {\nu,t} \right){\bf{D}}\left( \nu \right){\bf{E}}\left( {\nu,t} \right){\bf{K}}\left( {\nu,t} \right)
{\bf{F}}\left( {\nu,t} \right){\bf{Z}}\left( {\nu,t} \right)
}
with the parameters being defined as described in this section. 

\subsubsection{Additive Noise}

Understanding the noise characteristics is very important for the LOFAR EoR-KSP, which aims to detect very 
weak radio emission. The noise properties of the instrument are vital for a sensible error analysis. The total 
effective collecting area for the LOFAR-EoR experiment is $\sim0.07~{\rm km^2}$ at $150~{\rm MHz}$. The 
instantaneous bandwidth of the LOFAR telescope is $32~{\rm MHz}$ and
the aim for the LOFAR-EoR experiment is to observe in the frequency
range between 115 and 180$~{\rm MHz}$, which is twice the instantaneous
bandwidth. To overcome this, multiplexing in time has to be used
\citep[for more details see,][]{PP07}. For the purpose of this simulation
we ignore this complication and assume 400 hours of integration time on a single field
for a bandwidth of 32~\MHz, centred at $\nu_0=150$ \MHz.  This is chosen for two reasons: first, the frequency of 150~\MHz is the frequency where reionization presumably peaks in the current cosmological simulations. 
Second, due to the size of the data generated and the constraints of our current hardware (a 16-way 
symmetric multiprocessor system with 32 GB of memory), we are required to reduce the number of channels. 
This deteriorates the foreground fitting efficiency, but this is a reasonable compromise to be made. As we will 
show in the next section, this issue does not seem to pose any significant threat to our ability to statistically 
detect the EoR signal and our approach can thus be regarded as conservative.

The ultimate sensitivity of a receiving system is determined
principally by the system noise. The noise
properties of an elaborate receiving system like LOFAR can be very complex and we will address this more in 
forthcoming papers. The
theoretical $rms$ noise level in terms of the real and imaginary part of the complex visibilities for an interferometer 
pair between stations $p$ and $q$ is given by :
$$\Delta V\left\{ {{\mathop{\rm Re}\nolimits} ,{\mathop{\rm Im}\nolimits} } \right\}_{p,q}  = \frac{1}{{\eta _s }} \times 
\sqrt {\frac{{\mathrm{SEFD}_p  \times \mathrm{SEFD}_q }}{{2 \times \Delta \nu  \times \tau _{avg} }}} $$
where $\eta _s$ is the system efficiency that accounts for electronic,
digital losses, $\Delta \nu$ is the
frequency bandwidth and $\tau_{{\mathop{\rm{avg}}} }$ is the averaging time during which each station accumulates 
data.  SEFD is the System Equivalent Flux Density. The SEFD can be written as $\mathrm{SEFD} = T_{\mathrm{sys}} /
K$, where $K = (\eta _{\mathrm{a}}  \times A_{\mathrm{eff}})/(2 \times k_{\mathrm{B}} )$ and depends on the station efficiency 
$\eta_{\mathrm{a}}$ and the effective collecting area $A_{\mathrm{eff}}$ of the station. $k_\mathrm{B}$ is the Boltzmann constant.  For the 
system noise we assume two contributions. The first comes from the sky and is frequency dependent ($\approx 
\nu^{-2.55}$) and the second comes from the receivers. For the LOFAR HBA stations (24 tiles) the SEFD is  $\sim 
2000~{\rm Jy}$ at $150~{\rm MHz}$, depending on the final design \citep{PP07}. We assume that the SEFD 
varies within one per cent between different  stations. In order to calculate the SEFD we use the following system 
temperature ($T_{\mathrm{sys}}$) scaling relation as function of frequency ($\nu$): $$T_{\mathrm{sys}}=140+60(\nu/300~
\rm{MHz})^{-2.55}.$$ For our simulation, we scale the noise, so that in one night of observations we have the 
same sensitivity as the LOFAR-EoR observations after 400 hours. This is done because we currently lack the 
computational capacity to simulate and analyse a full data-cube of order several TBs per field. The equation above gives 
the noise on the real and imaginary parts of the visibility. When the SNR is high the noise distribution of the visibility 
amplitude and phase is Gaussian to an extremely good approximation. Without any signal the amplitude noise 
follows the Rayleigh distribution. When the SNR is low the measured amplitude noise follows the Rice distribution 
\citep{tms,taylor99,lopez} and gives a biased estimate of the true underlying signal.  For the 
phases, the noise distribution becomes uniform. 

In the following section we  describe one method that demonstrates our ability to statistically
detect the EoR signal from the post-calibration LOFAR maps that include realistic
levels for the noise and zero-mean calibration residuals. In both cases for the statistical detection of
the signal we use the total intensity maps only. In contrast to previous work, however, because we have 
taken polarization into account, its effects leaking into $Q, U$ and $V$ maps and are therefore accounted 
in our analysis.
Gain calibration is assumed to be done with a precision of two per cent, which is a realistic estimate, judging from 
the experience gained thus far from the LOFAR Core Station One (CS1) data analysis.

\subsection{The data--model inversion method}

To invert the ME, we have rewritten it as a linear equation, relating the observed visibilities to the true underlying 
visibilities, using the properties of the outer Kronecker product \citep{golub,boonstra05}. The relevant Jones--Mueller matrix describes the parameters 
of the model. This way we have to deal with a linear (over the parameters) model. Each 
parameter is a non-linear, multi-variate function of other parameters that describe the physics behind each of the 
effects described by the relevant Jones matrices. Due to our lack of information, though, we shall consider the 
elements of the Jones--Mueller matrix as multilinear functions of those parameters. Our task is to infer the 
parameters of the Jones--Mueller matrix, that we sample in the presence of Gaussian calibration errors and white noise.

The maximum-likelihood (ML) is a powerful method for finding the free parameters of the model to provide a good fit. 
The method was pioneered by R. A. Fisher \citep{fisher22} .  The maximum likelihood estimator (MLE) selects 
the parameter value which gives the observed data the largest possible probability density in the absence of a prior, 
although the latter can be easily incorporated. For small numbers of samples, the bias of maximum likelihood 
estimators can be substantial, but for fairly weak regularity conditions it can be considered asymptotically optimal 
\citep{mackay03}. With large numbers of data points, such as in the case of the LOFAR EoR KSP,  the bias of the 
method tends to zero. In general, it is not feasible to estimate the size of the data needed in order to obtain a 
good enough degree of approximation of the likelihood function to a multivariate Gaussian. 
We assume that errors associated with each datum are independent, but they have different variances and 
correlation properties over time and/or frequency.  Many of these problems can be overcome by MCMC (Markov Chain Monte Carlo) or 
{nested sampling} of the posterior \citep{skilling04,hobson08}.

Traditionally, the CLEAN/self-calibration approach has been used in radio-interferometry to give estimates of both the model parameters and the sky brightness distribution \citep{selfcal1}. Given an initial model of the sky distribution, which is usually made 
using CLEAN components for a few bright sources with a known structure, we solve for the non-linear parameters $\bf{p}$ of the 
model, and then improve the initial map. This is done in a loop until the optimization converges. In our approach, 
we make a distinction between the calibration and the data-model inversion parts. This is justified by the volume 
of the data to be handled. Another factor which is relevant for the EoR KSP is that we have to work near and 
below the noise level, and initial estimates of the sky distribution are going to be affected by the noise
and calibration errors.  We therefore need 
to make a proper determination of the true underlying visibilities and the model parameters, while retaining the 
noise and cosmological signal properties.

In our approach we assume that the parameter vector $\bf{p}$ is known within an accuracy of one to two per 
cent. For example, in Fig. \ref{fig:gain}, the gain used in the simulation and the recovered solution are shown. The parameter vector $\bf{p}$ includes all the (non)-linear parameters of the instrumental and ionospheric 
models, that are part of the Jones matrices (see section 4.2). This is a key assumption and its impact needs to be determined. We 
will address this issue in the future, when the LOFAR roll-out will provide us with more data to test our current 
error estimation levels. Given that, the maximum likelihood solution can be formulated in a more powerful matrix 
formalism, as:
\[
\left[ {{\bf{M}}^{\dag}_{j} \left( {\bf{p}} \right){\bf{N}^{-1}_{\rm noise}} {\bf{M}}_i\left( {\bf{p}} \right)} \right]{\bf{b}}_{\nu} = 
\left[ {{\bf{M}}^{\dag}_{j} \left( {\bf{p}} \right){\bf{N}^{-1}_{\rm noise}} } \right]{\bf{v}}_{\rm obs,\nu}, 
\]
where $\bf{N}_{noise}$ is the determinable noise covariances and $\bf{M}=\textbf{A}\otimes\textbf{A}$. Its inverse plays the role of a metric in the 
N-dimensional vector space of the data. The metric is useful for answering questions having to do with the 
geometry of a vector space. In our case we use it to actually compute the dot product on that space. 
Solving this equation gives the ML solution for Gaussian noise.  By linearizing the above ML equation along the 
eigenvectors of {\bf p}, we expect to find a better solution. Eventhough this can be done fully analytically, the 
complexity of the resulting matrices is so large that we plan to determine the later through forthcoming numerical simulations. By 
perturbing $\bf{p}$, we can then recover the proper noise and EoR covariance matrices and determine our final 
solution of {\bf s}. This process could even be iterated, but we expect convergence after a couple of iterations, since 
we start with an already good approximation of the model parameters.
With the addition of extra regularization terms, a deconvolved image can be obtained, from which the foreground 
and point-sources can be subtracted or 
filtered, to leave only the EoR signal and the noise. This `ideal' process, will be the subject of the second paper in 
this series. In the current paper, however, our main goal is to show that if calibration errors can be corrected without bias (i.e. errors on the correction have asymptotically zero mean), that we can indeed recover the EoR signal. Clearly, 
this is the first step in this process. In the following section we will therefore apply the more simple method used 
by \cite{jelic08}, which does not do a ML inversion, but simply uses ``dirty'' images with identical synthesized 
beam shapes, in order to extract the cosmological signal from the dirty maps.

\section{Results from the Signal Extraction}

The ultimate benchmark whether we can, in principle, recover the redshifted 21-cm signal from the 
LOFAR EoR KSP data set, is to apply our ML inversion and our signal-recovery procedure to the 
simulated data with realistic sky, ionospheric and instrument settings, and if that is successful, 
ultimately to real data sets. 
 
This, however, is not yet computationally feasible\footnote{Current tests indicate 30 hrs of
computing time for the ML inversion of a single frequency channel of realistic data size
on a single high-end CPU} and we opt in this paper for 
a less computationally expensive approach by removing the foregrounds from the ``dirty'' images. We are in fact implementing the full ML method on a mini-cluster of three
quad-core PCs connected to three NVIDIA  Graphics Processor Units.

\begin{figure*}
\centering
\begin{tabular}{cc}
\epsfig{file=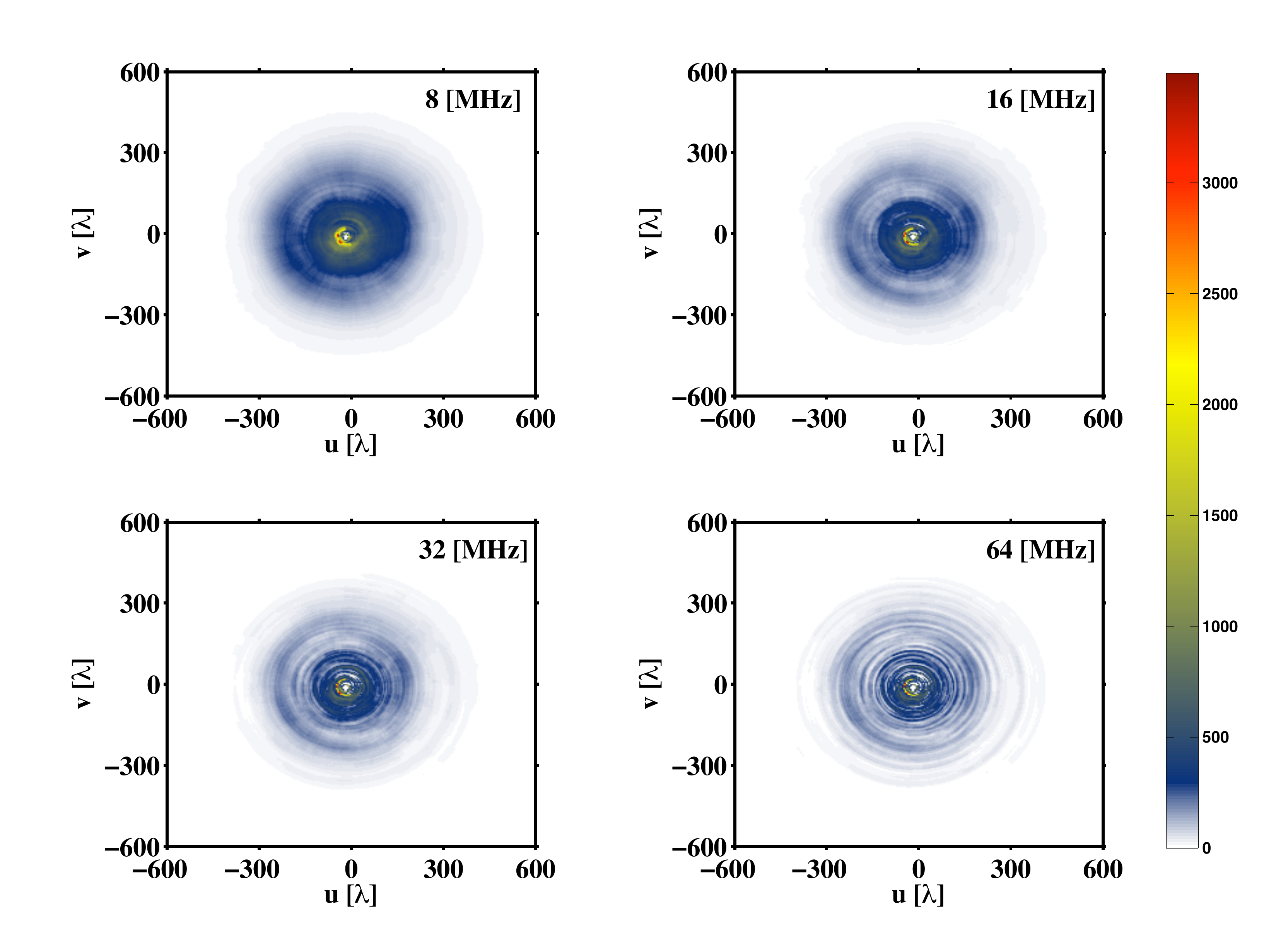,width=0.5\linewidth,clip=} &
\epsfig{file=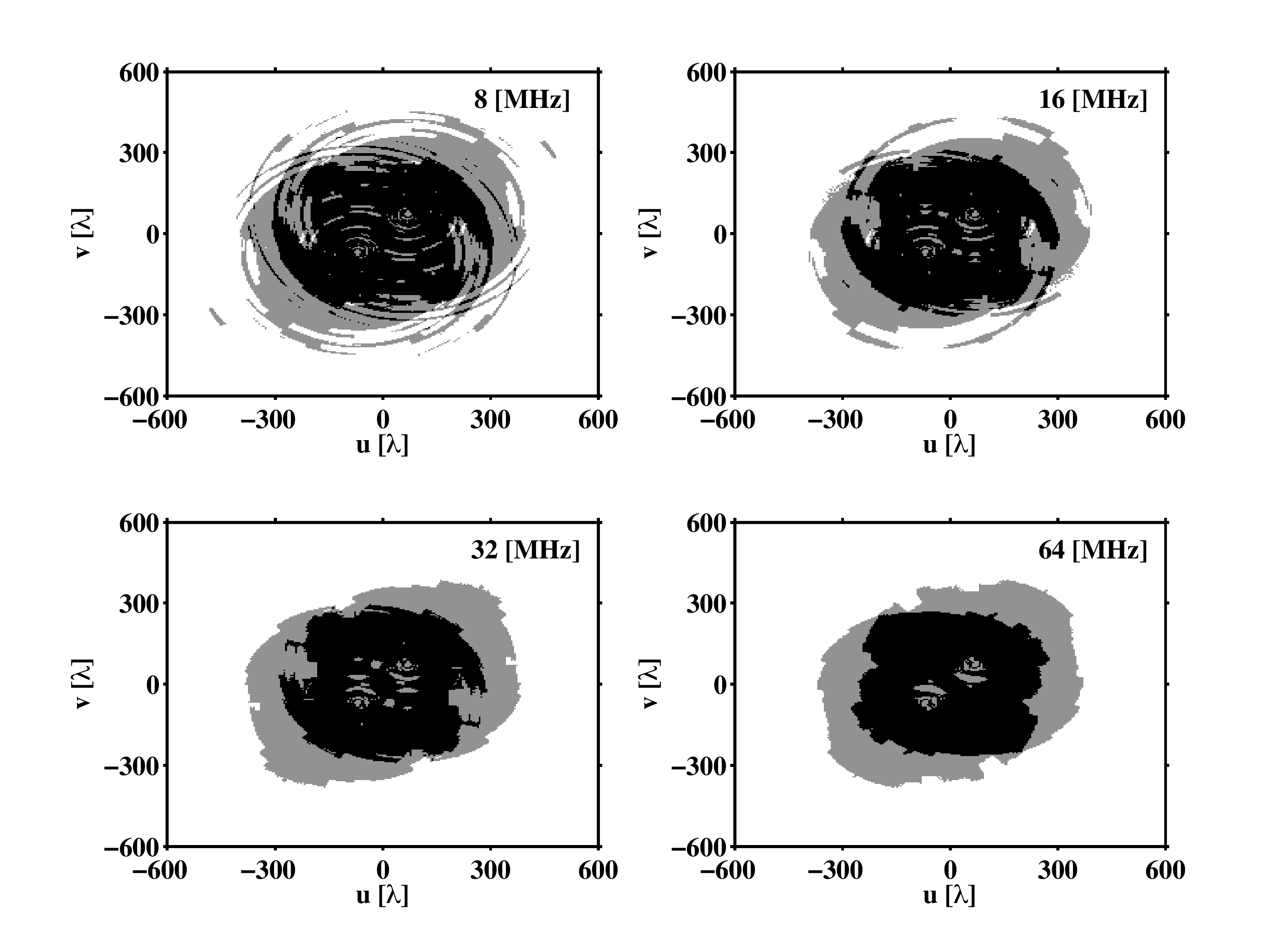,width=0.5\linewidth,clip=} 
\end{tabular}
\caption{The sampling of the uv-plane by the LOFAR core along frequency after 6 hours of synthesis. The left-hand 
figures show the average number of visibilities per uv cell for 8, 16, 32 and 64 MHz of total 
bandwith (the instantaneous bandwith of LOFAR is 32MHz). We assume that the data are delivered at 0.1 MHz 
resolution. The colorbar shows the number of visibilities per grid point. The right-hand set of figures shows the area in which less than 5$\%$ of the data along frequency is lost due 
to the scaling of the uv coverage with frequency, compared to the the total bandwidth. The black points represent 
regions where the visibilities and their Fourier conjugates occupy the same place, while the grey points represent 
true visibility measurements. This distinction is made because the Fourier conjugates do not contribute to the 
SNR.}
\label{fig:uvmask}
\end{figure*}

To remove the foregrounds, we currently use a 
polynomial fit along frequency for every line of sight on the ``dirty'' images. To produce the dirty maps we use only uv points that are present in 95\% of the frequency channels (Figure \ref{fig:uvmask}). This leads to an identical PSF for every map, with the additional cost of resolution loss. The visibilities are gridded on a super-sampled, regular grid of 256 by 256 cells.  This  degrades the resolution of the dirty maps somewhat but minimizes any
mixture between spatial and frequency fluctuations because of gaps in the uv-plane. Our current analysis is also done including only the Galactic foregrounds, and we will address the issue of the extragalactic 
foregrounds in the future, once we have more computational power available to us. 

An inappropriate polynomial fitting, however,
could remove part of the EoR signal or in the case of
under fitting of the foregrounds, fitting residuals could dominate
over the EoR signal (for details and discussion see \cite{jelic08}).
Hence, after substracting the foregrounds from the data cubes, the
residuals should ideally contain only the noise plus the EoR signal. The noise level is nearly an order of magnitude larger than the EoR signal, however,
so one is able to make only a statistical detection of the signal over the map by taking the
difference between the variance of the residuals and the variance of the
noise. The underlying assumption here is that the general statistical properties of
the noise are known to a high accuracy as a function of frequency. 
We are currently investigating methods that 
provide accurate noise estimates from the or partly calibrated data.

\begin{figure*}
\centering
 
		\includegraphics[width=0.8\textwidth]{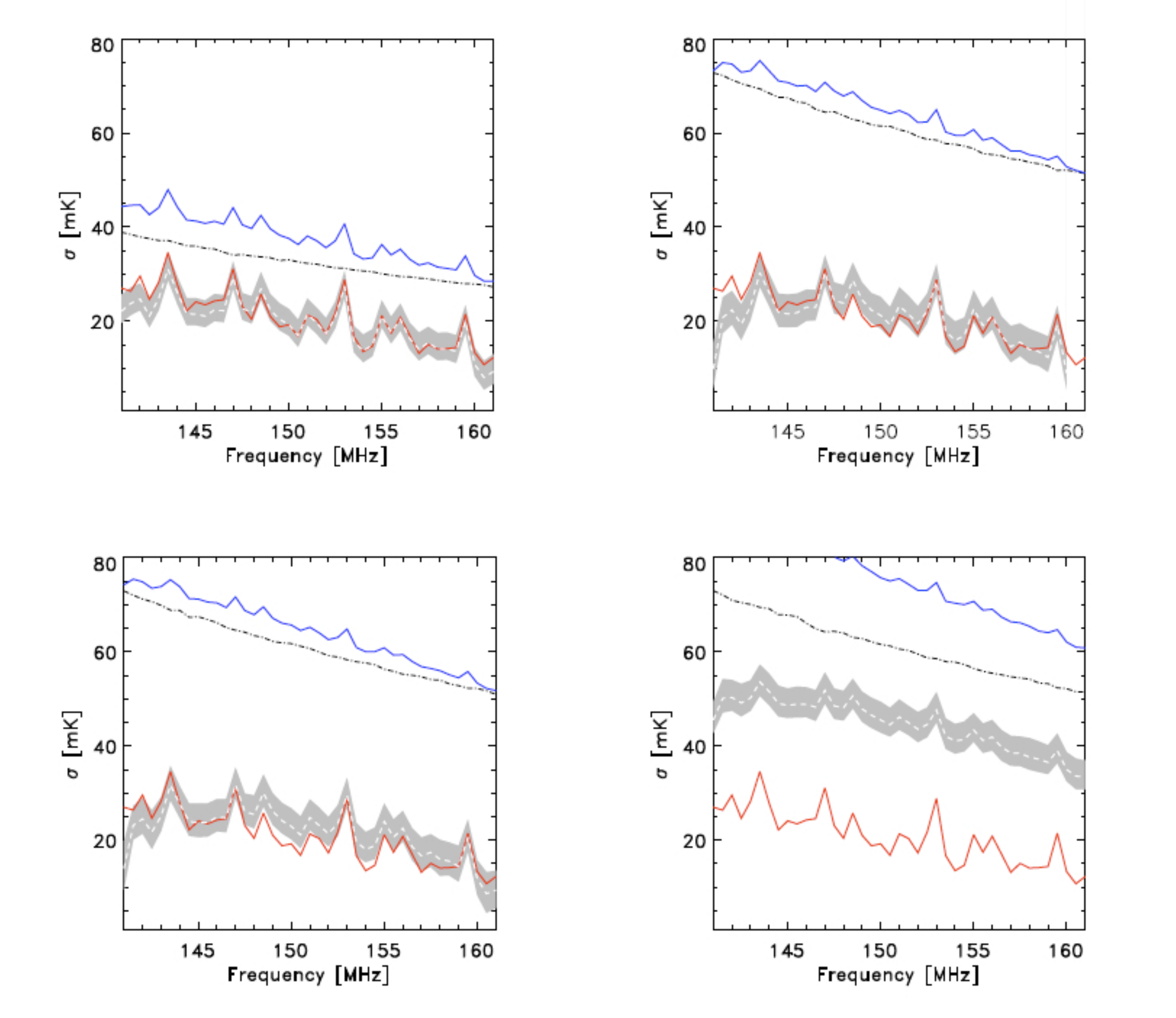} 
		
		\caption{The signal extraction for (a) 0.5 per cent (top left), (b) 1 percent (top right) , (c) 2 per cent  (bottom left) and (d) 10 per cent  (bottom right) errors on the calibration 
parameters. The solid blue line represents the $rms$ value of the residuals of the fiting procedure. The black-dotted line is the $rms$ of the instrumental noise as a function of frequency. The $rms$ of the ``dirty'' map of the 
cosmological signal is plotted with the red line and of the extracted signal with the white dashed line. The shaded 
grey areas are the errors on the $rms$ of the extracted signal calculated from 100 realizations of the noise. The 
effectiveness of the extraction proceedure decreases rapidly with the amplitude of the errors. 
}
\label{fig:signal_fit}
\end{figure*}

Fig. \ref{fig:signal_fit} shows the standard deviation of residuals as a function of
frequency (dashed line), after taking out the smooth component of the
foregrounds using a third-order polynomial. The white dashed line
represents the mean of the detected EoR signal after 100
independent Monte Carlo simulations of the extraction method applied
to each realization, while the grey shaded zone shows $2\sigma$ detection limits.
As one can see the detected EoR signal is in good agreement with the
original (solid red line) for most the frequencies (Table \ref{models}). We use bold-face letters to mark our current expectations. We thus conclude that in this case we can still recover the EoR signal, while for larger errors we gradually lose our ability to statistically detect the EoR signal.

\begin{table}
  \centering
  \caption{Values of the
  noise ($\sigma_{noise} [{\rm mK}]$ at $150~{\rm MHz}$), for different levels of calibration errors}
  \begin{tabular}{@{}ccccc@{}}
    \hline & case (a) & case (b) & case (c) & case (d) \\
    \hline $\rm{Error} \ [{\%}]$& 0.5 & \textbf{1} & 2 & 10\\
    $\sigma_{\rm{noise}} [{\rm mK}]$&  36 & \textbf{58} & 60 & 73\\
    \hline
  \end{tabular}\label{models}
\end{table}

\section{Summary \& Outlook}

The main goal of this paper has been to introduce a physics-based data model for the LOFAR EoR Key Science
Project, use this data model to generate realistic data sets, including Galactic and extragalactic foregrounds and 
known instrumental, ionospheric and noise properties (with zero-mean residuals), and subsequently 
extract the redshifted 21-cm EoR signal from the data cubes. Despite simplifications, which 
we indicated and will further address in forthcoming papers, we have clearly made a large step forward in 
simulating {\sl realistic} circumstances under which currently planned EoR experiments like LOFAR will
operate. This has thus far been lacking in many publications on this topic, which have all neglected
calibration and uv-sampling issues, and clearly needs to be addressed
beyond overly simplified assumptions before results from any of the currently planned projects
can be believed. 

We have also shown that using the currently known or estimated instrument specifications 
and layout of LOFAR, the
21-cm EoR signal can be recovered if the noise properties are known a priori (although we have not
yet tested whether the extraction of noise and EoR signal separately is possible), and the calibration 
of the ionosphere and instrument can be carried out in an unbiased fashion to the level that WSRT and LOFAR test setups
indicate is possible. In forthcoming papers we will address far more complex situations that include
all possible foregrounds (not only Galactic) and where calibration will be done using the simulated 
data themselves. This requires a considerable increase in our computational power and we are currently
implementing our codes on GPUs to achieve the required processing power. 

In more detail, in this paper we initially started by providing a brief description of the physical connections behind the Hamaker--Bregman--Sault 
formalism. This completes the picture established by \cite{hbs1}. The connection to physics helps in better 
understanding new and old problems in radio-polarimetry and  provides a new perspective. 
Novel methods of  calibration and image evaluation can be developed based on those principles
and we indicated a few possibilities (see Appendix A). 

After describing the connection between the physics of the Jones and Mueller calculi and the data model for a 
generic radio interferometric array, we introduced the data model that is relevant for the LOFAR
EoR KSP. This 
description of the signal pathways was used to generate detailed, large-scale polarized instrumental response 
simulations for several instrumental parameters assuming simplified, but still realistic models for their values and 
errors. The high-resolution data produced by this simulation were generated in such a way that they resemble a typical LOFAR EoR observation. 

We applied the same procedure, albeit simple, as described by \cite{jelic08} to recover the 
cosmological signal, and as a first benchmark of the inversion method. The polynomial fit method 
\citep{zaldarriaga04,jelic08} gives a reasonable result, even in the presence of small errors in the calibration 
parameters. When including realistic instrumental noise and calibration errors, we can still recover the 
cosmological signal if the properties of instrumental noise are well known as a function of frequency. In order to 
obtain the frequency dependence of the noise, we used the true underlying sky distribution. In reality, one might  
not have this luxury. 
However, differencing of image or visibility values between narrow frequency channels could in principle deliver the noise properties of the instrument as function of frequency.

What has not been addressed in this paper, but will be in the follow-up paper, is the calibration
process itself. The traditional self-calibration algorithm \citep{selfcal1} alternates between calibration and CLEAN steps in order to find the optimal 
image, from a given visibility data set. This is done by substracting simple models for the sky brighness distribution 
and correcting for the calibration solutions at every step. This, however, leads to a suboptimal solution and certainly cannot 
deal easily with complex sky brightness structures \citep{starck02,sanjay04,ent1,ent2,ent3,kembal}. 
On the other hand, the maximum likelihood method has traditionally been used in 
many disciplines in order to estimate the parameters of  linear systems. It can address the calibration, 
(after linearization), and 
deconvolution problems and in forthcoming papers we will further test this methodology, besides
others [e.g. Expectation Maximization; \cite{sarod08b}].

Finally, there is the problem of the signal extraction itself. Although done independently of the 
inversion process in this paper, we suspect it can possibly be done simultaneously with the inversion.
This will also be addressed in forthcoming papers.
Here, we have studied signal extraction through polynomial fitting, but matched filtering is 
another approach that we plan to take, and we suspect it might be a good complementary method. Those algorithms might have to make some use of a priori assumptions, though, since the observational 
evidence for the cosmological signal is scarce to non-existent.

A detailed analysis of the noise properties of the foreground-cleaned maps also still has to be carried out. This can be done in a Monte Carlo 
sense. This way one can study the asymptotic efficiency of the ML, using the observed information matrix. 
Another consideration should be the correlation properties between the instrumental parameters, as well as the 
error propagation, since in principle their behaviour is highly non-linear. 

Despite a considerable increase in the complexity of large-scale EoR simulations (Thomas et al. 2008), foregrounds models (Jelic et al. 2008) and the instrument model (this paper) -- necessary to process 
and understand the results from our forthcoming LOFAR EoR KSP observations -- with this paper
we intend to provide a guide towards even more complex simulations, including all the effects that we mentioned 
above and thus far have avoided or implemented in a simplified manner. 
We conclude with the comment,  learned from the current paper, that only by doing 
simulations and successful signal-recovery on a scale
and complexity-level comparable to the real observations, can one convincingly show 
that ongoing experiments --
be it LOFAR, MWA, or otherwise -- are capable of extracting the redshifted 21-cm signal from
data sets affected by noise, RFI, calibration errors and bright foregrounds. It is also essential
in interpreting their ultimate results.  

\bigskip

\section*{Acknowledgments}

The authors would like to thank Johan Hamaker, Jan Noordam, Oleg Smirnov and Stephan Wijnholds for useful discussions 
during the various stages of this work. LOFAR is being funded by the European Union, European Regional
Development Fund, and by ``Samenwerkingsverband Noord-Nederland'', 
EZ/KOMPAS.

\appendix

\section{Properties of the Stokes Vector and Matrix}
In this appendix we introduce several fundamental concepts in the description of polarized radiation. Although not yet used in this paper, we feel that they are important to mention and will be used in forthcoming publications to guide us to a better understanding of complex calibration processes in radio interferometry.

\subsection{Pauli Matrices}

Pauli matrices are well known and have been used for the analysis of partially polarized light \citep{fano53}. Their major 
advantage is that they satisfy a set of properties that significantly reduce the complexity of calculations 
associated with the intensity. The identity plus the Pauli matrices in two dimensions are defined as
\e{\begin{array}{*{20}c}
   {{\bf{\sigma }}_0  = \left( {\begin{array}{*{20}c}
   1 & 0  \\
   0 & 1  \\
\end{array}} \right)} & {{\bf{\sigma }}_1  = \left( {\begin{array}{*{20}c}
   1 & 0  \\
   0 & { - 1}  \\
\end{array}} \right)}  \\~\\
   {{\bf{\sigma }}_2  = \left( {\begin{array}{*{20}c}
   0 & 1  \\
   1 & 0  \\
\end{array}} \right)} & {{\bf{\sigma }}_3  = \left( {\begin{array}{*{20}c}
   0 & { - i}  \\
   i & 0  \\
\end{array}} \right)}  \\
\end{array}}
This set of $2 \times 2$ linearly independent  matrices constitute a basis for the vector space of $2 \times 2$ 
Hermitian matrices over the compex numbers. Summarizing their properties, they are Hermitian and 
they follow the commutation relations $\left\lfloor {{\bf{\sigma }}_i ,{\bf{\sigma }}_j } \right\rfloor  = {\bf{\sigma }}_i 
{\bf{\sigma }}_j  - {\bf{\sigma }}_j {\bf{\sigma }}_i  = i2\varepsilon _{ijk} {\bf{\sigma }}_k $
 where $\varepsilon _{ijk}$ is the Levi-Civita permutation symbol \citep{golub,fano53,boonstra05}. These matrices are unitary and traceless 
except for the identity matrix. The linear expansion of the coherency matrix in this basis is
 \e{
 	{\bf{C}} = \frac{1}{2}\sum {{\rm tr}\left( {{\bf{C\sigma }}_i } \right)} {\bf{\sigma }}_i
	}
with $s_i  = {\rm tr}\left( {{\bf{C\sigma }}_i } \right)$ being the four Stokes parameters: $i= 0,1,2,3$ corresponding 
to the Stokes $I$, $Q$, $U$ and $V$, respectively. 

\subsection{The Stokes Vector and Matrix}

In the literature the Stokes parameters are usually arranged as a $4 \times 1$ vector,  ${\bf s} =(I,Q,U,V)^{T}$ 
\citep{gil07}. An alternative notation would be to introduce a $2 \times 2$ Stokes matrix.  
\begin{equation}
	{\bf S} = \frac{1}{2}\left(\begin{array}{*{20}cc}
	I + Q & U -i V \\
	U + i V & I-Q
		\end{array}\right)
\end{equation}
In this case for pure states we have $\left\| {\bf{C}} \right\|^2  = \frac{1}{2}\left\| {\bf{S}} \right\|^2 $
 and in the case of unpolarized light $\left\| {\bf{C}} \right\|^2  = \frac{1}{4}\left\| {\bf{S}} \right\|^2 $. This is because 
for unpolarized light $Q$, $U$ and $V$ are zero and the matrix becomes $I$ multiplied by the identity matrix. Since there is a 
factor of $\frac{1}{2}$, the Frobenius norm of the matrix has this factor squared. Even if the source is not 
polarized one should observe the same signal in both orthogonal polarizations, as in this case both polarization 
states are equiprobable. This is an important fact for calibration. The X and Y dipoles of the LOFAR HBAs should 
measure the same signal, for unpolarized sources, and any fluctuations should be due to polarization calibration 
errors.

\subsection{Polarization Level} 

In addition, the degree of polarization is defined as $P \equiv  \sqrt{U^2  + V^2  + Q^2 } /{I}$. 
We can also define the unit vector \citep{gil07} 
$$
	{{\mathbf{p} }}^T  = \frac{1}{P I}
	\left( \begin{array}{*{20}c}
		 Q \\ 
		 U \\
		 V 
	\end{array} \right)
$$ 
 The Stokes vector can be decomposed using those parameters in several ways. A trivial decomposition is 
between a polarized and unpolarized state. As radio receivers are intrinsically polarized, and in the case of 
LOFAR orthogonal, a spectral decomposition can express the Stokes vector as a convex linear sum of two 
orthogonal pure states. This makes the relationship between the measured signals and the parameters 
describing the system more clear, since we have to deal with orthogonal states. The spectral decomposition of 
the coherency matrix is using its eigenvalue structure to decompose it into pure states, i.e.\ eigenvectors. The 
spectral decomposition (diagonalization) of the coherency matrix is then equivalent to
\[
{\mathbf{s}} = I \times \left( {\frac{{1 + P}}
{2}\left[ {\begin{array}{*{20}c}
   1  \\
   {\mathbf{p }^{T}}  \\

 \end{array} } \right] + \frac{{1 - P}}
{2}\left[ {\begin{array}{*{20}c}
   1  \\
   {\mathbf{-p}^{T}}  \\

 \end{array} } \right]} \right)
\]

Of course, for a mixed state there are infinite combinations of independent states into which it can be decomposed, 
but it would be useful if the selection is such that it matches the characteristics of the system. The parameters $I
$ and $P$ are invariant under unitary transformations (changes of coordinate systems). They are directly related 
to the eigenvalues of $\bf C$. Pure states are related to rank-1 polarization matrices, and mixed states to 
rank-2. Wolf showed that there always exist two orthogonal reference directions, such that the degree of 
coherence is maximized and coincides with $P$. This is important because various random distributions 
correspond to unpolarized light. The measurement of the correlations of the Stokes parameters allows us to 
distinguish between those different types of non-polarized light.

\subsection{Entropy} Concluding with the quality criteria, the von Neumann entropy can be applied to 
electromagnetic waves \citep{fano57,brosseau}.  In terms of the coherency matrix  it is defined as $S =  - {\rm tr}\left( {{\mathbf{\hat C}}\ln 
{\mathbf{\hat C}}} \right)$. Is a measure of the difference in the amount of information between pure and mixed 
states, both with same intensity. It can be expressed as a function of the eigenvalues of $\bf C$ as $S =  - 
\frac{1}
{2}\left[ {\left( {1 + P} \right)\ln \left( {\sqrt {1 + P} } \right) + \left( {1 - P} \right)\ln \left( {\sqrt {1 - P} } \right)} \right]$. 
It is a decreasing monotonic, bounded function of $P$ \citep{gil07}. It attains its maximum value, $S=\ln(2)$, for non polarized light 
and its minimum, $S=0$, for totally polarized light. Using this entropy one can define the polarization 
temperature. Depolarizing effects, such as the ionospheric Faraday rotation, can be studied this way in order to 
detect spatial heterogeneity. For example, ionospheric TIDs should lead to an increase in the polarization 
entropy. This is a good scheme of ranking observations: if the polarization entropy differs between two maps, 
residual Faraday rotation and leakage can be present. Finally, for non-Gaussian distributions of polarization 
states, higher order moments are needed and the Stokes system is no longer adequate to describe the 
polarization states.

\begin{figure*}
	\centering
 \vspace{-30mm}
		\includegraphics[width=1.1\textwidth]{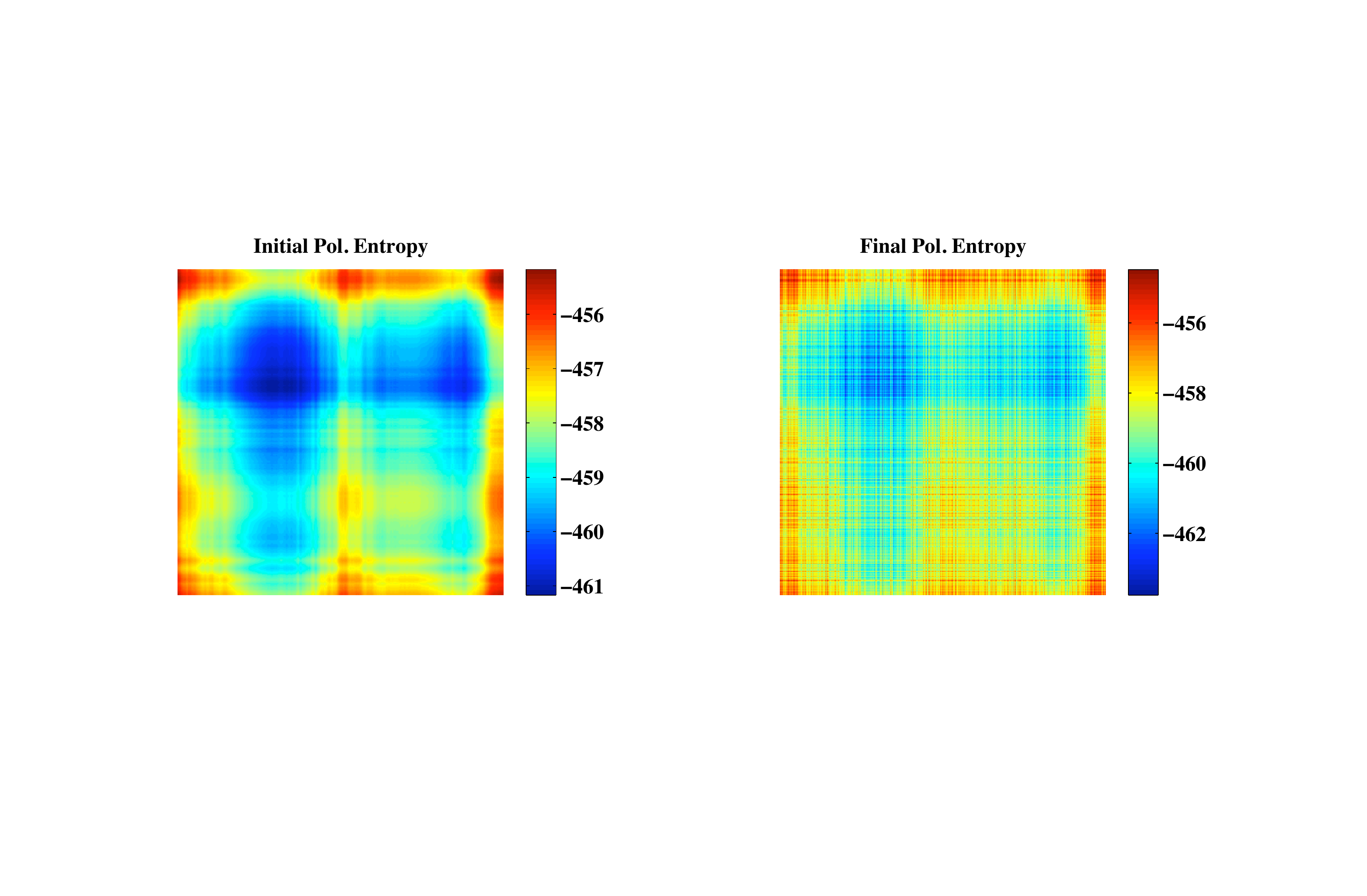} 
\vspace{-40mm}
	\caption{Left: the polarization entropy of a simulated Galactic diffuse synchotron emission map. Right: the same plot after applying a depolarizing effect (Faraday rotation) along the horizontal axis}
	\label{fig:entropy}

\end{figure*}

The standard (self)-calibration procedure for interferometric observation of polarized light aims at recovering the 
coherency matrix of the polarized radiation. Given the statistical nature of the coherency matrix, we must 
emphasize the importance of parameters that give a measurement of their polarimetric purity. Polarization 
entropy is a concept related to the impurities of the media through which radiation propagates, as it was defined 
above. It is useful for many purposes, especially when depolarization is a relevant subject. An increase in the 
polarization entropy signifies a decrease in the polarization purity. It is a direct way to access the quality of the 
data. Ionospheric Faraday rotation causes a depolarization with a certain frequency behavior. We would expect 
that the entropy would follow this behavior. 
In Figure \ref{fig:entropy} we show the polarization entropy for several lines of sight as a function of frequency in a simulated map. Attention should be brought to the fact that is really hard to distinguish between Faraday rotation and other depolarization effects.

\subsection{Lorentz Transformations}
The transformations of the Stokes vector lead to a system of differential equations \citep{landi}. 
From the statistical interpretation of the coherency matrix we derive that the Stokes $I$ parameter has to be positive 
and that $s_0  = I \ge s_1^2  + s_2^2  + s_3^2  = Q^2  + U^2  + V^2 $, which means that there can be no more 
polarized light than the total light. This is a key concept in physics, namely the conservation of energy. Extending this remark, we can define a 
Minkowski space (in the mathematical sense) for the Stokes 4-vectors. The norm in this space would be $\left\| 
{\mathbf{s}} \right\| = I^2  - Q^2  + U^2  + V^2 $. Positive values of the norm correspond to the time-like vectors of 
the special theory of relativity. Light-like Stokes vectors correspond to totally polarized states. The Poincare 
sphere defines the light cone with the exception that the symmetric part of the cone does not have any special 
meaning. The Jones vector transforms as usual under the Lorentz group. The coherency matrix is defined 
through the Kronecker product of two Jones vectors and is a 2$\times$2 spinor of rank two. The Stokes vector and the 
coherency matrix must be realizations of the same irreducible representation of the Lorentz group, as they both 
have 4 independent components. The Lorentz transformations form a 6 parameter group. The six generators are 
the 3 spatial rotation generators of O(3) $\bf R$ and the Lorentz boosts $\bf B$. The infinitesimal transformation 
of the Stokes vector is \[
{\mathbf{s}}'  = {\mathbf{s}} - \sum\limits_i^3 {\left( {r_i {\mathbf{R}}_i  + b_i {\mathbf{B}}_i } \right)} {\mathbf{s}}
\operatorname{d} I = {\mathbf{s}} - {\mathbf{Ks}}\operatorname{d} I
\]
where $\bf K$ is a matrix that resembles the absorption matrix. This equation is the radiative transfer equation 
along the signal path. If one choses to exclude effects that intensify the radiation, this is the complete mathematical 
description of the effects in the signal path. The Lorentz boosts describe polarizing effects, while the spatial 
rotations describe the Faraday rotation. The relation between the initial and final Stokes vector is a finite Lorentz 
transformation. While it sounds straightforward it is not an easy task. The problem arises from the non 
commutativity of the generators. Magnus has proposed a solution to this problem. We will discuss this soon.

\subsection{Clifford Algebra} 

After having introduced the coherency matrix and the Jones formalism we proceed a step further in the 
mathematical abstraction in order to unify all possible cases. We note that every system is equivalent to a 
parallel combination of pure systems, and any pure system to a combination of retarders and diattenuators. In 
\citet{hbs4} the author proposes the use of the quaternion algebra to describe the Jones matrices, or their 
$SO^{+}(1,3)$ covering group. Quaternions, as any other space isomorphic to $\mathbb{C}$ constitute a subalgebra of a Clifford algebra. In particular, in three dimensions the Clifford product is equivalent to spatial rotations. 
Pure systems can still be described as Lorentz transformations in this algebra. Since algebraic manipulations 
become more clear, the computational requirements of problems involving partial polarization can be reduced. The 
Clifford algebra of three-dimensional space $Cl_3$ represents the four dimensional Minkowski space time. Clifford algebras have been used extensively to describe polarization mode dispersion (PMD) in optical fibres \citep{reimer}. In that field they have to deal with a spatially inhomogeneous, birefringent medium (the optical fibre), which resembles the effects caused by the ionosphere on the EM waves that propagate through it.

\subsection{Magnus Expansion} In the case of polarization mode dispersion, which can occur due to ionospheric 
Faraday rotation, transversal of the polarized source signal through a set of phase screens etc, one is 
interested in recovering the original polarization state of the radiation. Two possibly quasi-orthogonal modes, 
represented by the input frequency-independent and the output Jones vectors, are related through a complex 
$2\times2$ Jones matrix. Despite the fact that we might not have any prior knowledge about the structure of the 
medium through which radiation propagates, the frequency dependence of that effect contains useful information 
about that medium. For example the $\sim \lambda ^2$ Faraday rotation should leave a distinct imprint on the 
signal, which can help us distinguish it from other depolarizing effects. In the general case, the coherency matrix 
is transformed as ${\mathbf{C}}^{\mathbf{'}} {\mathbf{ = JCJ}}^{\mathbf{\dag }} $. An arbitrary Jones matrix with 
determinant 1, can be also expressed in terms of two vectors $\bf{a}$
 and $\bf{b}$ according to \e{{\mathbf{J}} = \exp \left[ { - \left( {{i \mathord{\left/
 {\vphantom {i 2}} \right.
 \kern-\nulldelimiterspace} 2}} \right)\left( {{\mathbf{b}} + i{\mathbf{a}}} \right) \cdot \sigma } \right],}\\ 
where $\bf{\sigma}$ is the Pauli spin vector. This is a Jones matrix representation of the Lorentz transformation in a 
Clifford algebra. In the case of a frequency dependent effect, 
${\mathbf{J}}\left( \omega  \right)$, the Jones space operator can be decomposed into real and imaginary 
components as \e{\begin{gathered}
  {\mathbf{J}}\left( \omega  \right) = \exp \left[ { - \left( {{i \mathord{\left/
 {\vphantom {i 2}} \right.
 \kern-\nulldelimiterspace} 2}} \right)\left( {{\mathbf{b}} + i{\mathbf{a}}} \right) \cdot \sigma } \right], \hfill \\
  {\mathbf{\tilde J}}_{\mathbf{\omega }} {\mathbf{\tilde J}} =  - \frac{i}
{2}\left[ {{\mathbf{\Omega }} + i{\mathbf{\Lambda }}} \right], \hfill \\ 
\end{gathered} } where the subscript $\omega$ represents differentiation with respect to the frequency. A 
solution for $\bf{J}$ can be obtained through the Magnus expanion, which specifies 
${\mathbf{J}}\left( \omega  \right) = \exp \left( {\sum\limits_{n = 0}^\infty  {{\mathbf{B}}_n \left( \omega  \right)} } 
\right){\mathbf{J}}\left( {\omega _0 } \right)$, where the first two coefficients are given by \[
\begin{array}{*{20}c}
   {{\mathbf{B}}_1 \left( \omega  \right) = \int_{\omega _0 }^\omega  {d\omega _1 J_\omega  \left( {\omega _1 } 
\right),} }  \\
   {{\mathbf{B}}_2 \left( \omega  \right) = \int_{\omega _0 }^\omega  {\int_{\omega _0 }^{\omega _1 } {d\omega _2 
d\omega _1 } \left( {J_\omega  \left( {\omega _1 } \right)J_\omega  \left( {\omega _2 } \right) - J_\omega  
\left( {\omega _2 } \right)J_\omega  \left( {\omega _1 } \right)} \right)} }.  \\

 \end{array} 
\]

The coefficients of the Magnus expansion for $n>2$, are related to those of lower order through recursion. Taylor 
expanding the frequency derivatives of the Jones matrix to third order gives us a way to directly evaluate the 
Magnus coefficients. To the extent of our knowledge, the methods mentioned above have not been applied in the field of radio interferometry.



\label{lastpage}

\end{document}